\documentclass[aps,pra,groupedaddress,nofootinbib,notitlepage,showpacs,floatfix,twocolumn,
https://www.overleaf.com/project/5bc975e487db8c5f557c1e70 amsmath,amssymb]{revtex4-1}

\usepackage{graphicx}% Include figure files
\usepackage{dcolumn}% Align table columns on decimal point
\usepackage{bm}% bold math
\usepackage{physics}
\usepackage{subfigure}
\graphicspath{{./figures/}}
\usepackage{color}
\usepackage{hyperref}
\usepackage{url}
\hypersetup{
    colorlinks=true, 
    linkcolor=cyan,
    citecolor=magenta, 
    filecolor=magenta, 
    urlcolor=cyan,
    runcolor=cyan
}

\begin{document}
% -------------- author and title ------------------

\title{Dynamics of reverse annealing for the fully-connected $p$-spin model}

\author{Yu Yamashiro}
\affiliation{Department of Physics, Tokyo Institute of Technology,  Nagatsuta-cho, Midori-ku, Yokohama 226-8503, Japan}
\affiliation{Jij Inc.,
High tech Hongo Building 1F, 5-25-18 Hongo, Bunkyo, Tokyo 113-0033, Japan}

\author{Masaki Ohkuwa}
\affiliation{NTT DATA Mathematical Systems Inc.,
Shinanomachi, Shinjuku-ku, Tokyo, 160-0016, Japan}

\author{Hidetoshi Nishimori}
\affiliation{Institute of Innovative Research, Tokyo Institute of Technology, Nagatsuta-cho, Midori-ku, Yokohama 226-8503, Japan}
\affiliation{Graduate School of Information Sciences, Tohoku University, Sendai 980-8579, Japan}

\author{Daniel A. Lidar}
\affiliation{Departments of Electrical Engineering, Chemistry, and Physics \& Astronomy, University of Southern California,
Los Angeles, California 90089, USA}
\affiliation{Center for Quantum Information Science \& Technology, University of Southern California, Los Angeles, California 90089, USA}

\begin{abstract}
Reverse annealing is a relatively new variant of quantum annealing, in which one starts from a classical state and increases and then decreases the amplitude of the transverse field, in the hope of finding a better classical state than the initial state for a given optimization problem.  We numerically study the unitary quantum dynamics of reverse annealing for the mean-field-type $p$-spin model and show that the results are consistent with the predictions of equilibrium statistical mechanics. In particular, we corroborate the equilibrium analysis prediction that reverse annealing provides an exponential speedup over conventional quantum annealing in terms of solving the $p$-spin model. This lends support to the expectation that equilibrium analyses are effective at revealing essential aspects of the dynamics of quantum annealing. We also compare the results of quantum dynamics with the corresponding classical dynamics, to reveal their similarities and differences. We distinguish between two reverse annealing protocols we call adiabatic and iterated reverse annealing. We further show that iterated reverse annealing, as has been realized in the D-Wave device, is ineffective in the case of the $p$-spin model, but note that a recently-introduced protocol (``$h$-gain"), which implements adiabatic reverse annealing, may lead to improved performance.
\end{abstract}

\maketitle
% -------------- author and title ------------------

\section{Introduction}
Quantum annealing (QA) is a quantum-mechanical metaheuristic for combinatorial optimization problems \cite{kadowaki_quantum_1998, brooke_quantum_1999, santoro_theory_2002, santoro_optimization_2006, das_colloquium:_2008, morita_mathematical_2008}, and its strict adiabatic realization is known as adiabatic quantum computation  \cite{farhi_quantum_2000, farhi_quantum_2001,albash_adiabatic_2018}. In the conventional formulation of QA, the initial condition is chosen to be the ground state of the transverse field, which is a uniform superposition of all possible classical states. Perdomo-Ortiz \textit{et al}. proposed some time ago \cite{perdomo-ortiz_study_2011} to instead start from an appropriately chosen classical state and gradually increase and then decrease the amplitude of the transverse field to find a better solution than the initial classical state. They called this protocol `sombrero adiabatic quantum computation' based on the shape of the amplitude of the transverse field, 
%which is now named reverse annealing, 
and showed by means of a few numerical examples that the method is indeed effective if the initial condition is properly chosen. Further refinements were proposed in Ref.~\cite{Chancellor:2016ys}, and the protocol whereby the initial state is classical rather than a quantum superposition is now called reverse annealing. The current generation of the D-Wave quantum annealer, D-Wave 2000Q, implements reverse annealing, and several recent studies have been reported on using this feature in different contexts, e.g., quantum simulation \cite{king_observation_2018}, matrix factorization \cite{Ottaviani2018}, portfolio optimization \cite{Venturelli2018}, and mid-anneal pausing \cite{Marshall2019}.

A theoretical study of reverse annealing was initiated recently by some of us in Ref.~\cite{ohkuwa_reverse_2018}, in which equilibrium statistical mechanics was employed to study when and how reverse annealing is effective. It was shown explicitly that reverse annealing can turn first order quantum phase transitions into second order transitions in a simple problem, the $p$-spin model, by choosing an appropriate annealing path.  It is generally the case that a first-order quantum phase transition is associated with an exponentially closing energy gap $\Delta$ between the ground state and the first excited state as a function of system size, whereas a second order transition is associated with a polynomially closing gap (though exceptions are known~\cite{Pfeuty1970,Laumann:2012hs,Tsuda2013}). On the other hand, according to the adiabatic theorem of quantum mechanics, the computation time $\tau$ is inversely proportional to a small power of the energy gap $\Delta$ \cite{Kato:50,Jansen:07,lidar:102106}. Taken together, these facts suggest that reverse annealing provides an exponential speedup relative to the conventional QA protocol, for the $p$-spin model.
In this work we study this intriguing connection from the dynamical perspective, by numerically solving the time-dependent Schr\"odinger equation of the reverse annealing protocol for the $p$-spin model. Our goal is to test whether the exponential speedup suggested by the static, equilibrium statistical mechanics analysis does indeed hold.

This is non-trivial since, while a quantum annealer at very long annealing times is likely to experience a quasistatic evolution, returning a final population that is close to a Boltzmann distribution of the Hamiltonian at a single (freeze-out) point during the annealing~\cite{Amin:2015qf}, results from equilibrium statistical mechanics clearly do not reveal all aspects of the dynamical behavior of quantum systems.
In particular, quantum annealing on relatively short timescales compared to the quantum gap can give rise to distinctly non-equilibrium features such as coherent oscillations due to quantum interference~\cite{Munoz-Bauza:2019aa,Karanikolas:2018aa}.

In the present work we evaluate the unitary, closed-system dynamics of reverse annealing by direct numerical integration of the time-dependent Schr\"odinger equation for large system sizes. This is possible due to a special symmetry of the $p$-spin model.  To this end, it is useful to distinguish two types  of reverse annealing. The first is the one adopted in Ref.~\cite{ohkuwa_reverse_2018} to facilitate analytical treatment by equilibrium statistical mechanics. In this formulation, an additional term is introduced into the Hamiltonian, which is usually composed of just two terms representing the classical cost function and the transverse field. The additional, third term serves to enforce a given classical ground state as an initial condition.  The system is then expected to follow an adiabatic path to the final state. We call this protocol \emph{adiabatic reverse annealing} (ARA).  In the second formulation the initial state is a classical state, an excited state of the initial (classical) Hamiltonian (the cost function), as formulated in Refs.~\cite{perdomo-ortiz_study_2011,Chancellor:2016ys}. The annealing process therefore follows a complicated combination of instantaneous eigenstates of the conventional two-term Hamiltonian, without any additional term.  It is clear that dynamical studies are indispensable for this case. This latter protocol is reverse annealing as is widely used in practice on the D-Wave device~\cite{king_observation_2018,Ottaviani2018,Venturelli2018,Marshall2019}, and 
is often used iteratively in practice, that is, the final state of a single cycle is fed into the system as its initial state of the next cycle under the expectation that the result improves iteratively. We therefore call this protocol \emph{iterated reverse annealing} (IRA).

This paper is organized as follows.  In Sec. \ref{Sec:ARA}, we analyze adiabatic reverse annealing. After reviewing the formulation and results on static properties of the $p$-spin model, we solve the time-dependent Schr\"odinger equation numerically for large systems to test the consistency of dynamics and static properties.  We also study two classical versions of the model to clarify the similarities and differences between quantum and classical dynamics.  Section \ref{Sec:iRA} deals with iterated reverse annealing by numerical solution of the time-dependent Schr\"odinger equation, as well as by spectral analysis of the instantaneous Hamiltonian with an emphasis on the special structure of the $p$-spin model.  The last section is devoted to conclusions.

%%%%%%%%%%%%%%%%%%%%

\section{Adiabatic reverse annealing}
\label{Sec:ARA}

In adiabatic reverse annealing (ARA), we add a term $\hat H_{\text{init}}$ to the Hamiltonian to fix the initial state to the ground state of the initial Hamiltonian \cite{ohkuwa_reverse_2018},
\begin{equation}
  \begin{split}
  \hat H(t) =&s(t)\hat H_0 + \qty(1-s(t))\qty(1-\lambda(t))\hat H_{\text{init}} \\
             &+ \Gamma \qty(1-s(t))\lambda(t) \hat V_{\text{TF}},
  \label{eq:ARA_H}
  \end{split}
\end{equation}
where $\hat H_0$ is the cost function to be minimized and is a function only of the 
%$z$ component of 
Pauli matrices $\{\hat\sigma_i^z\}$. The last term represents the transverse field,
\begin{align}
    \hat V_{\text{TF}} = -\sum_{i=1}^N \hat\sigma_i^x,
\end{align}
where $N$ is the total number of sites (qubits). The parameter $\Gamma$ controls the strength of the transverse field relative to the other terms. The initialization term is written as
\begin{align}
     \hat H_{\text{init}} &= -\sum_{i=1}^N \epsilon_i \hat\sigma^z_i,
\end{align}
where $\{\epsilon_i(=\pm 1)\}$ denotes the given classical initial state. This initial Hamiltonian fixes $\hat\sigma_i^z$ to $\epsilon_i$ as its ground state. The time-dependent parameters \(s(t)\) and \(\lambda(t)\) both change from 0 to 1 as time $t$ proceeds from 0 to $\tau$, where $\tau$ is the computation time. Thus the initial Hamiltonian is $\hat H_{\text{init}}$ and the final one is $\hat H_0$. The quantum term $\hat V_{\text{TF}}$ is effective only in intermediate times. The existence of the term \(\hat H_{\text{init}}\) allows us to analyze the system properties by equilibrium statistical mechanics under the assumption of adiabatic evolution.\footnote{
A recent update of the D-Wave quantum annealer introduced a function `$h$-gain', which can be used to realize ARA \cite{hgain}.}

We study the $p$-spin model in this paper,
\begin{align}
  \hat H_0 = -N\qty(\frac{1}{N}\sum_{i=1}^N \hat \sigma_i^z)^p,
  \label{eq:H_each_term}
\end{align}
where $p\geq 3$ is an integer.
%, three or larger. 
The ground state of this model is trivial, $\sigma_i^z=1~\forall i$ (and $\sigma_i^z=-1~\forall i$ for $p$ even). This simple problem is known to be hard for conventional quantum annealing 
%is known to have a difficulty to solve this simple problem 
due to a first order phase transition \cite{Jorg2010}. Since $\hat H_0$ and $\hat V_{\text{TF}}$ are symmetric with respect to an arbitrary permutation of the site index $i$, we do not lose generality by the following assignment of $\epsilon_i$,
\begin{equation}
  \epsilon_i = \begin{cases}
    +1 &\text{for}~ i \leq Nc\\
    -1 &\text{for}~ i > Nc,
  \end{cases}
\end{equation}
where the probability $c~(1/2\le c \le 1)$ for $\epsilon_i$ to be $+1$ is chosen such that $Nc$ is an integer. This means that the initial magnetization is \(2c-1\). Thus, the larger $c$ is, the closer the initial state is to the correct answer to the optimization problem, $\sigma_i^z=1,~\forall i$.\footnote{For $p$ even, another degenerate ground state $\sigma_i^z=-1~(\forall i)$ exists.  But we do not lose generality by restricting ourselves to the subspace with non-negative magnetization.}

Permutation symmetry allows us to write the Hamiltonian in terms of only a set of total spin operators of two subsystems,
\begin{equation}
 \hat S^{x,z}_1 \equiv \frac{1}{2}\sum_{i=1}^{Nc}\hat\sigma^{x,z}_i,\quad
 \hat S^{x,z}_2 \equiv \frac{1}{2}\sum_{i=Nc+1}^{N}\hat\sigma^{x,z}_i
\label{ra:eq:s1s2}
\end{equation}
as
 \begin{align}
 \hat H_0 &= -N\qty(\frac{2}{N}(\hat S^z_1 + \hat S^z_2))^p,\\
 \hat H_{\text{init}} &= -2(\hat S^z_1 - \hat S^z_2),~
 \hat V_{\text{TF}}=-(\hat S^x_1 + \hat S^x_2).
 \label{eq:H_blocked}
\end{align}
Since \((\hat{\bf{S}}_1)^2\) and \((\hat{\bf{S}}_2)^2\) commute with the total Hamiltonian, the system stays in the subspace with the largest values of \((\hat{\bf{S}}_1)^2\) and \((\hat{\bf{S}}_2)^2\) because the initial state lies in this subspace. This greatly reduces the size of the Hilbert space to be studied, from exponential to polynomial in $N$, \(\mathcal{O}(N^2)\).

\subsection{Static properties}

\begin{figure}
\centering
    \includegraphics[width=0.98\linewidth]{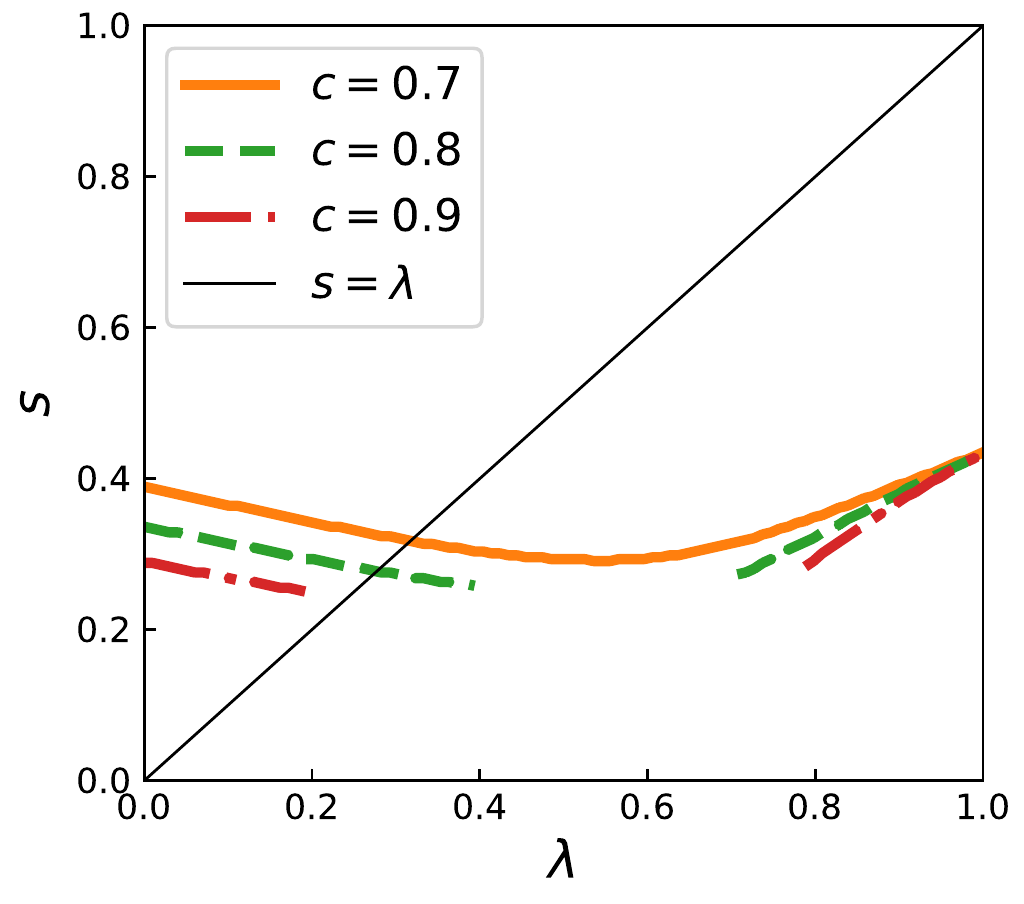}
  \caption{Static phase diagram  for $\Gamma=1$ and $p=3$. The axis $\lambda=1$ corresponds to the conventional QA. ARA starts from $\lambda=s=0$ and ends at $\lambda=s=1$. Lines for $c=0.7, 0.8$ and 0.9 represent first order phase transitions.
  }
      \label{fig:phase-diag_G1}
 % \label{fig:gap_with_phase}
\end{figure}

We first review the results presented in Ref.~\cite{ohkuwa_reverse_2018} with some generalizations. A statistical-mechanical analysis of the $p$-spin model for the case of ARA of Eq.~(\ref{eq:ARA_H}) with $s$ and $\lambda$ taken as fixed parameters leads to the following expression for the free energy at zero temperature (the ground state):
\begin{equation}
\begin{split}
 & f = s(p-1)m^{p-1}\\
  &- c \sqrt{\left[ spm^{p-1} + (1-s)(1-\lambda)\right]^2+\Gamma^2(1-s)^2 \lambda^2}\\
  &-( 1 - c )\sqrt{\left[spm^{p-1}-(1-s)(1-\lambda)\right]^2 + \Gamma^2(1-s )^2\lambda^2}~,
\end{split}
\end{equation}
where $m$ satisfies the following self-consistency condition to minimize the free energy:
\begin{equation}
  \begin{split}
  &m = c\frac {spm^{p-1}+(1-s)(1-\lambda)}
  {\sqrt{\left[spm^{p-1}+(1-s)(1-\lambda)\right]^2 +\Gamma^2(1-s)^2 \lambda^2}}\\
  &+(1-c)\frac{spm^{p-1}-(1-s)(1-\lambda)}
  {\sqrt{\left[spm^{p-1}-(1-s)(1-\lambda)\right]^2 + \Gamma^2(1-s)^2 \lambda^2}}.
  \end{split}
\end{equation}
Derivations are detailed in Appendix A of Ref.~\cite{ohkuwa_reverse_2018}, where $\Gamma$ was fixed to 1, but a generalization to arbitrary $\Gamma$ as given here is straightforward. We focus our analysis on the case of $p=3$, but other values of $p (\ge 4)$ show qualitatively the same behavior.
Figure~\ref{fig:phase-diag_G1} is the $\lambda$-$s$ phase diagram for $\Gamma=1$.
\begin{figure}
\centering
  \includegraphics[width=0.98\linewidth]{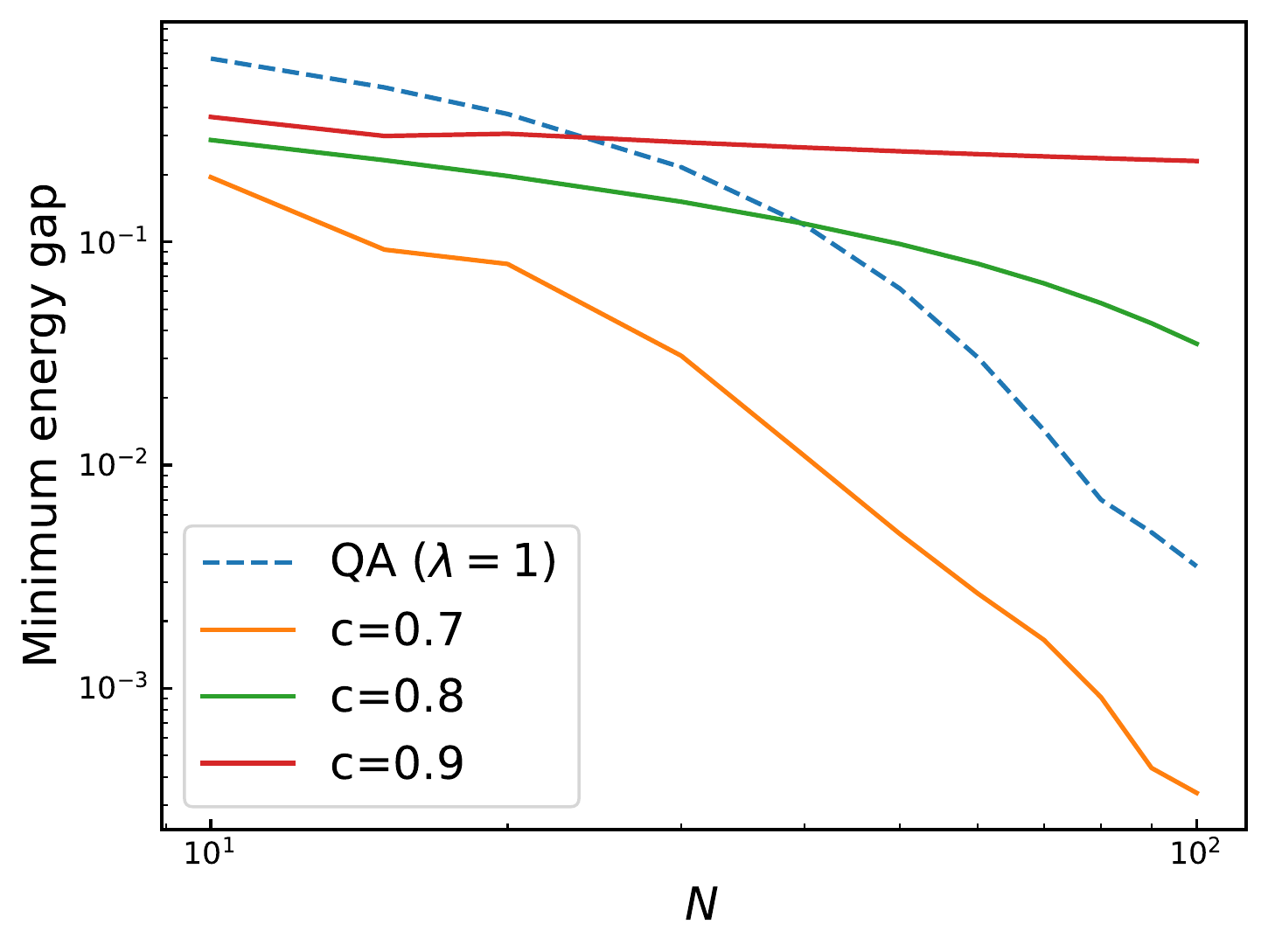}
  \caption{System size dependence of the energy gap for conventional QA (\(\lambda=1\)), shown dashed, and ARA (\(\lambda=s\)) for several values of \(c\). The linear behavior in the $c=0.9$ case is consistent with inverse polynomial scaling, while the curvature seen for $c=0.7$ and $0.8$ is consistent with inverse exponential scaling. Here, as in Fig.~\ref{fig:phase-diag_G1}, we set $\Gamma=1$ and $p=3$.}
    \label{fig:gap_vs_N}
\end{figure}
\begin{figure*}
\centering
  \subfigure[\(\Gamma=2, p=3\)]{
    \includegraphics[width=0.31\textwidth]{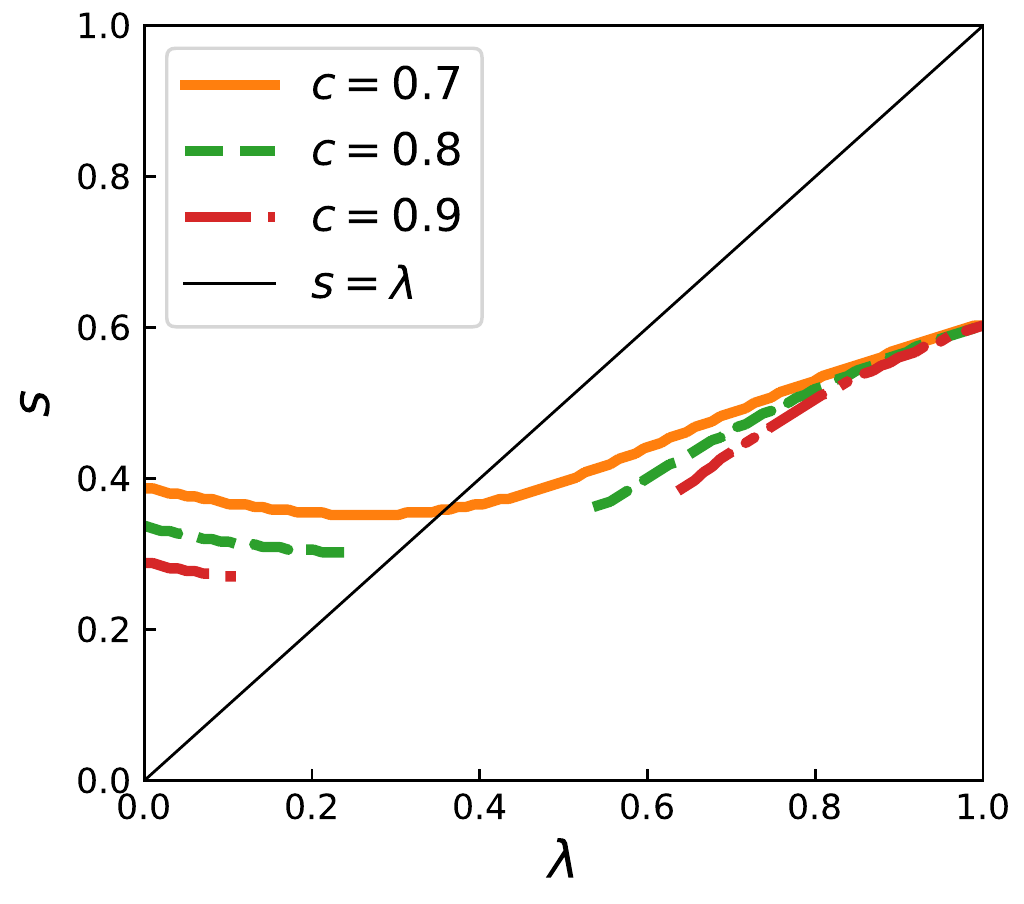}
    \label{fig:phase-diag_G2}
  }
  \subfigure[\(\Gamma=5, p=3\)]{
    \includegraphics[width=0.31\textwidth]{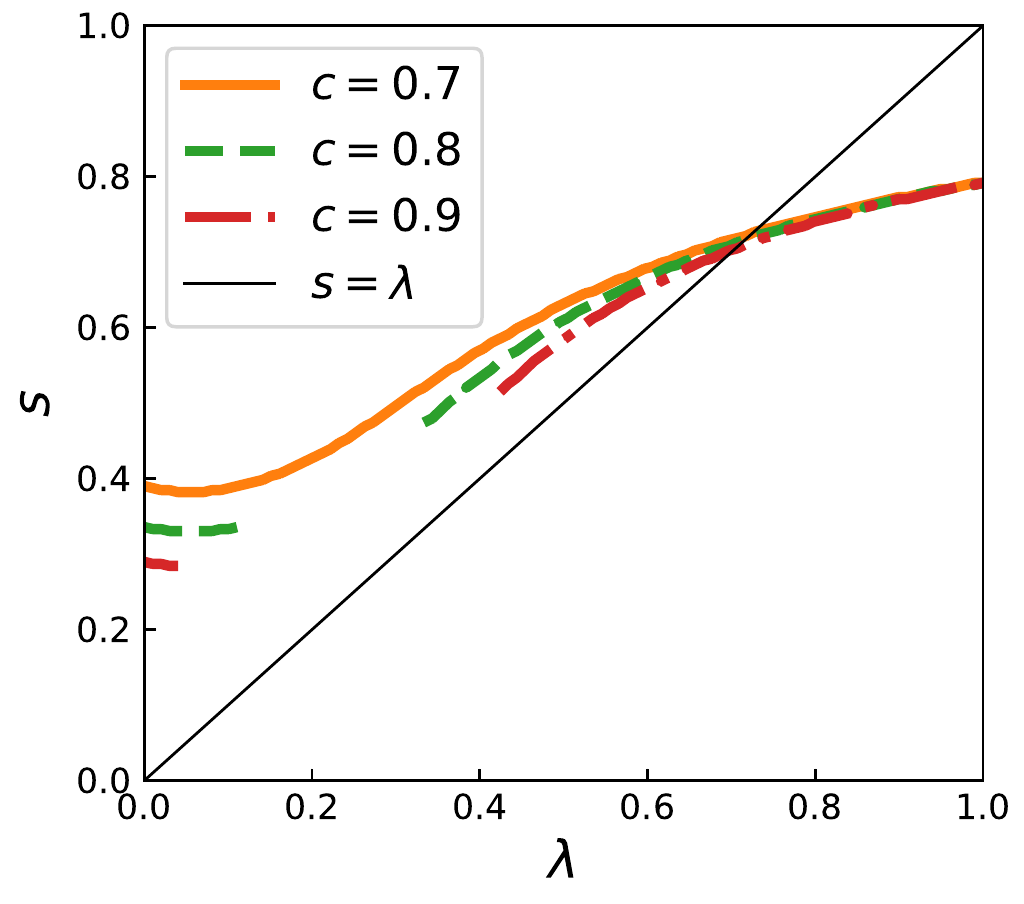}
    \label{fig:phase-diag_G5}
  }
  \subfigure[\(\Gamma=10, p=3\)]{
    \includegraphics[width=0.31\textwidth]{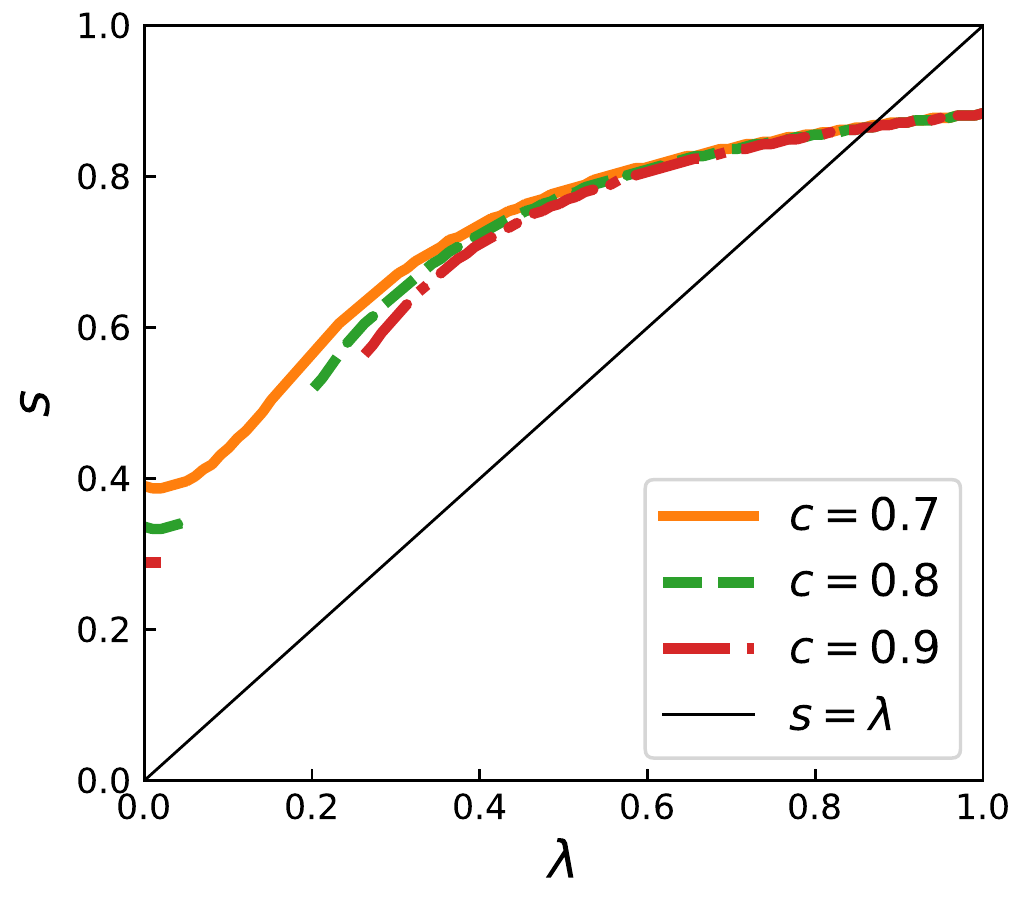}
    \label{fig:phase-diag_G10}
  }
  \caption{Phase diagrams in the \(s\)-\(\lambda\) plane for \(p=3\) for different values of the amplitude of transverse field. Curves indicate first order phase transitions.}
  \label{fig:phase_eachG}
\end{figure*}
The axis $\lambda=1$ corresponds to conventional QA, where $s$ is changed from 0 to 1. Along this axis, a first order phase transition at $s \simeq 0.4$ marks where QA becomes inefficient, since the energy gap closes exponentially there. As $\lambda$ is reduced from 1 toward the inner part of the phase diagram, the effects of the initial Hamiltonian $\hat H_{\text{init}}$ come into play, and the line of first order phase transitions is split into two parts, one for larger $\lambda$ and the other for smaller $\lambda$, if $c$ is above a threshold, i.e., if the initial condition is reasonably close to the final answer. This means that a path exists that connects the initial point $\lambda=s=0$ and the goal $\lambda=s=1$ without crossing a first-order transition. In particular, when we choose the path $s=\lambda$, the diagonal of the phase diagram, the system encounters a first order phase transition when $c=0.7$ and $0.8$ but not for $c=0.9$.  Since a larger $c$ means that the initial condition is closer to the correct answer, the phase diagram shows that a good choice of the initial classical state exponentially accelerates the computation by removing an exponentially closing energy gap.

The relation between the order of a phase transition predicted by statistical mechanics and the scaling of the energy gap for large but finite $N$ is not proven in general, and should be checked by other methods. Indeed, while usually a first (second) order transition is accompanied by an exponentially (polynomially) closing gap, as mentioned in Introduction there exist examples in which the gap closes non-exponentially at a first order phase transition~\cite{Pfeuty1970,Laumann:2012hs,Tsuda2013}. We therefore evaluated the energy gap by direct numerical diagonalization, which is possible for relatively large system sizes due to the high symmetry of the problem. As seen in Fig.~\ref{fig:gap_vs_N}, which is a log-log plot of the gap as a function of system size for $s=\lambda$,
only the data for $c=0.9$ behave non-exponentially. This conforms with the standard expectation from the phase diagram of Fig.~\ref{fig:phase-diag_G1}, where the path $s=\lambda$ avoids the line of first order transitions only when $c=0.9$. It is interesting that the data for $c=0.8$ behaves almost polynomially up to about $N=50$. This should reflect the fact that the line $s=\lambda$ crosses the first-order transition line close to its termination point in Fig.~\ref{fig:phase-diag_G1}: close to the termination point, which is a critical point or a second-order transition point, the width and height of the energy barrier between two coexisting states should be small, and the system's behavior is close to that at a second-order transition, as long as the size is not too large.

Larger values of $\Gamma$ change the phase diagram quantitatively, if not qualitatively. As seen in Figs.~\ref{fig:phase-diag_G2} ($\Gamma=2$),~\ref{fig:phase-diag_G5} ($\Gamma=5$), and~\ref{fig:phase-diag_G10} ($\Gamma=10$),
the line of first order transitions for $c=0.8$ and $0.9$ on the right part of the phase diagram extends toward the left part as $\Gamma$ increases. This is to be expected, because for large $\Gamma$ the intermediate Hamiltonian is mostly dominated by the transverse-field term $\hat V_{\text{TF}}$, and is close to the initial Hamiltonian of conventional QA.  It then follows that the first order transition of conventional QA persists even for relatively large values of $c$.

\subsection{Schr\"{o}dinger dynamics}

\begin{figure}[htpb]
  \centering
   \subfigure[\ \(\Gamma=1\)]{
    \includegraphics[width=0.9\columnwidth]{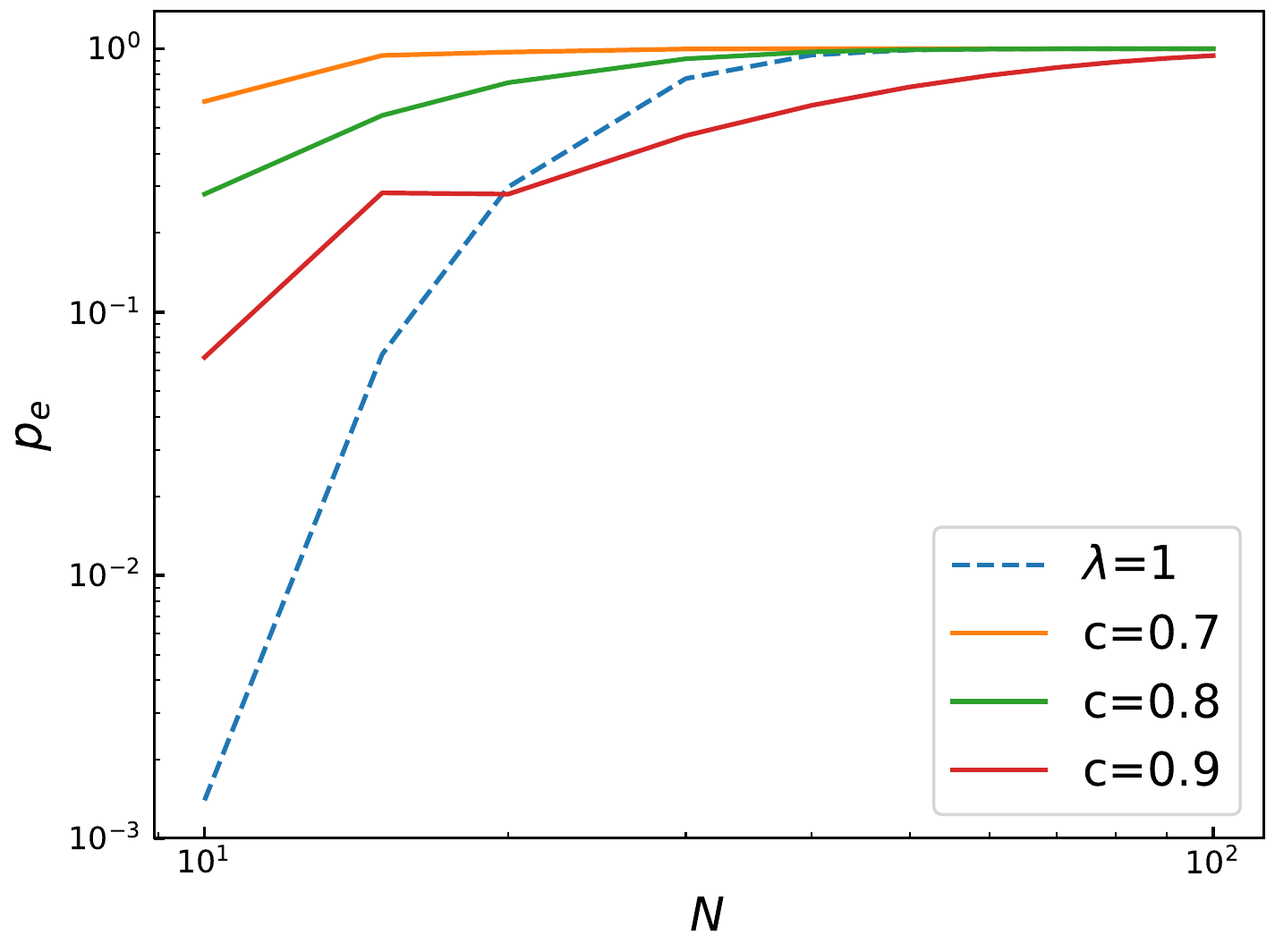}
  }
   \subfigure[\ \(\Gamma=2\)]{
    \includegraphics[width=0.9\columnwidth]{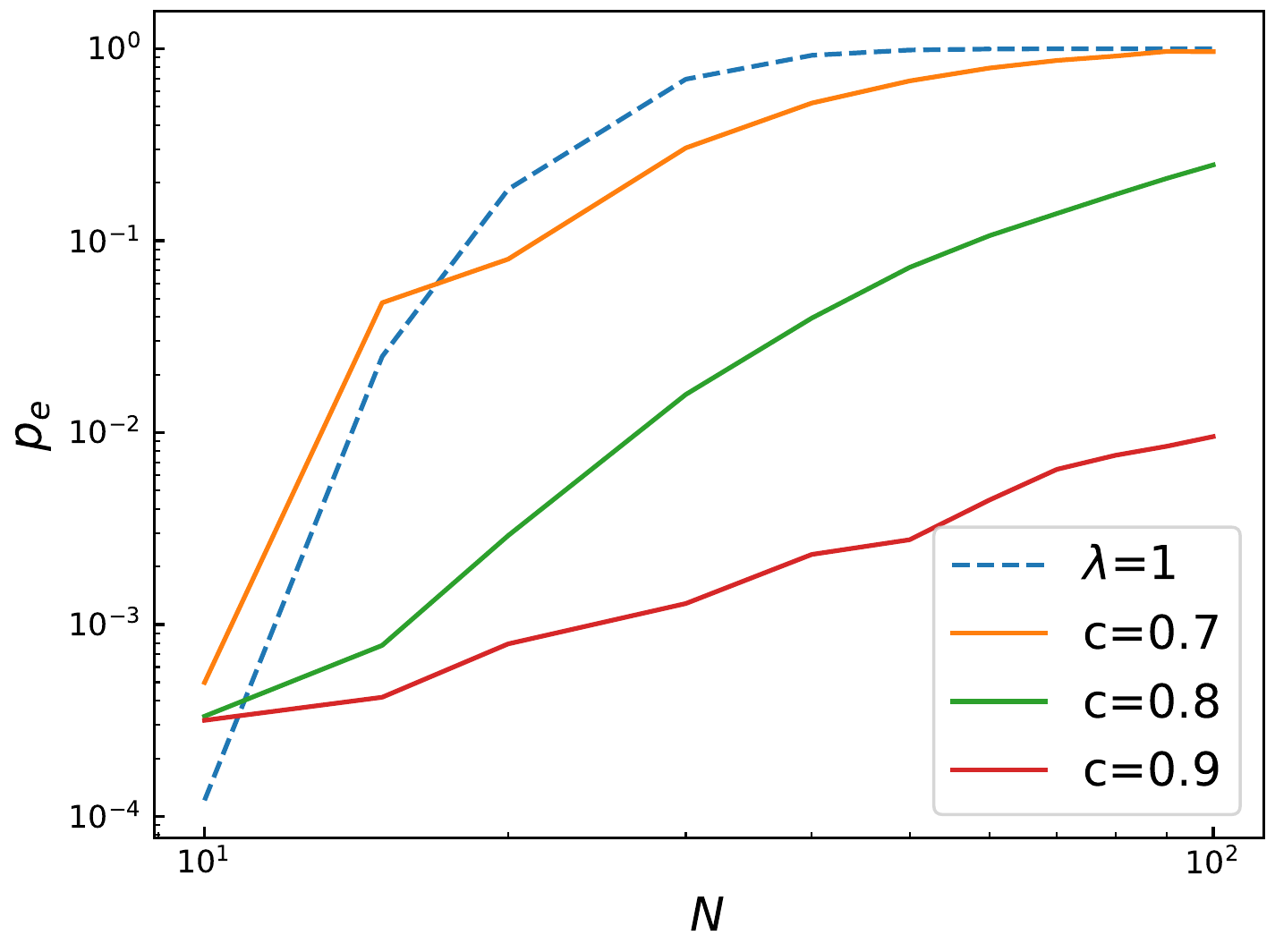}
  }
  \caption{
  Error probability as a function of system size for $p=3$ and $\tau=100$. In panel (a) we set \(\Gamma=1\) and in (b) \(\Gamma=2\). Blue dashed lines are for conventional QA.
  In panel (a), the final error probabilities of ARA are close to those of conventional QA. In panel (b), in contrast, ARA with $c=0.8$ and $0.9$ has much smaller errors than QA. }
  \label{fig:pe_vsN}
\end{figure}

We next report our results on the closed system dynamics. Let \(\ket{\phi}\) denote the ground state of the cost function \(\hat H_0\) and \(\ket{\psi(\tau)}\) the actual state reached after time $\tau$, subject to Schr\"{o}dinger dynamics. Then 
\begin{align}
  p_{\text{e}}(\tau) = 1 - \qty|\bra{\phi}\ket{\psi(\tau)}|^2
  \label{eq:dynamics_results2}
\end{align}
is the error probability, i.e., the probability of not reaching the ground state. This quantity  is plotted in Fig.~\ref{fig:pe_vsN} as a function of system size. For \(\Gamma=2\), the error probability increases polynomially for large $c$, which is consistent with the phase diagram in Fig.~\ref{fig:phase_eachG}.

A standard measure of computation time in quantum annealing is the time to solution (TTS), as defined by~\cite{speedup}
\begin{align}
%  p_{\text{e}}(\tau) = 1 - \qty|\bra{\phi}\ket{\psi(\tau)}|^2, \\
  \text{TTS}(\tau, p_d) = \tau \frac{\log(1-p_d)}{\log p_e(\tau)} .
  \label{eq:dynamics_results}
\end{align}
The TTS is the effective time it takes using ``runs" lasting time $\tau$ to find the correct solution at least once with probability $p_d$, which we set to $0.99$ or higher. 

\begin{figure}
  \centering
  \subfigure[\ $\Gamma =1$]{
    \includegraphics[width=0.9\columnwidth,pagebox=cropbox]{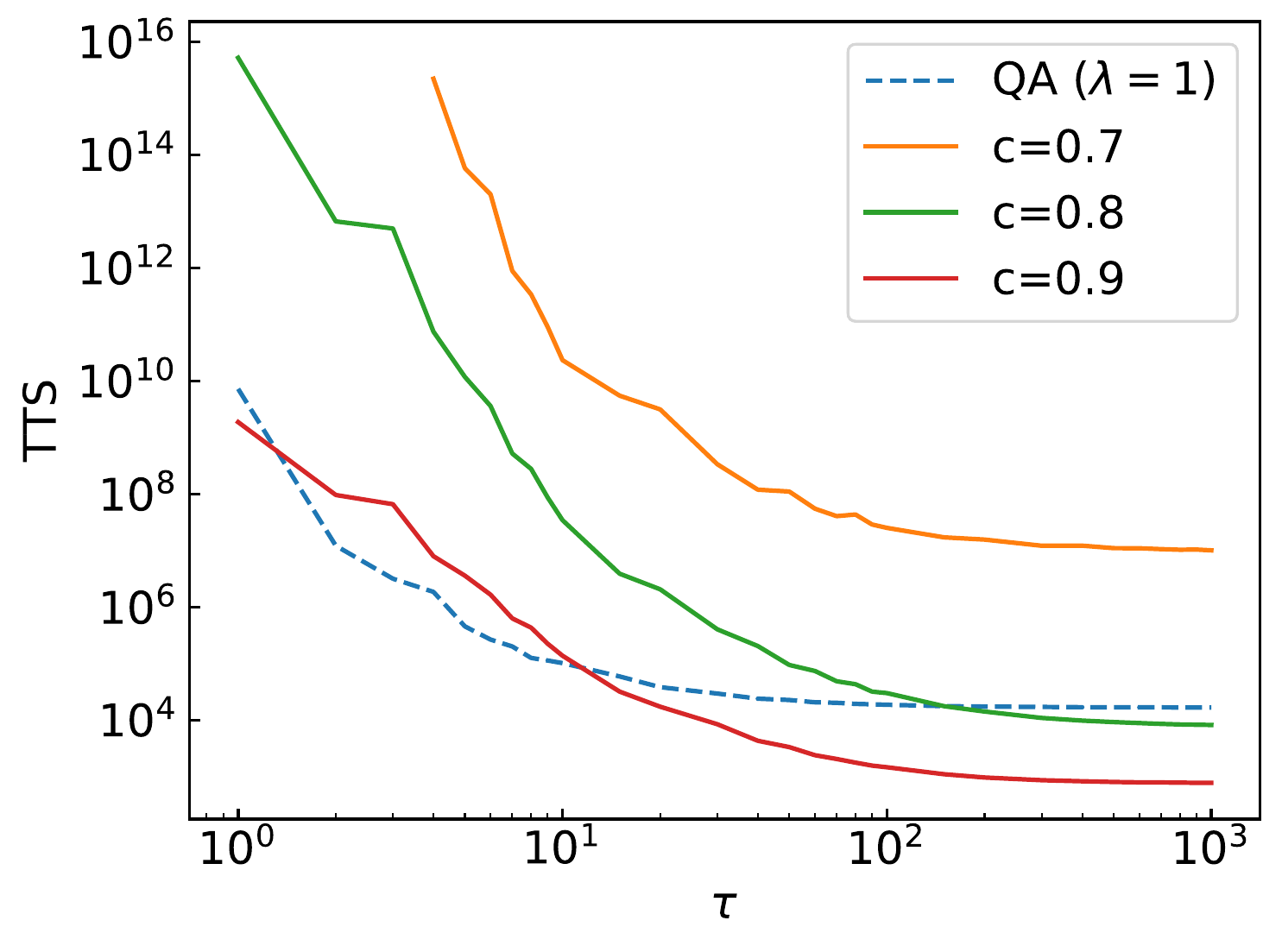}
  }
  \subfigure[\ $\Gamma=2$]{
    \includegraphics[width=0.9\columnwidth,pagebox=cropbox]{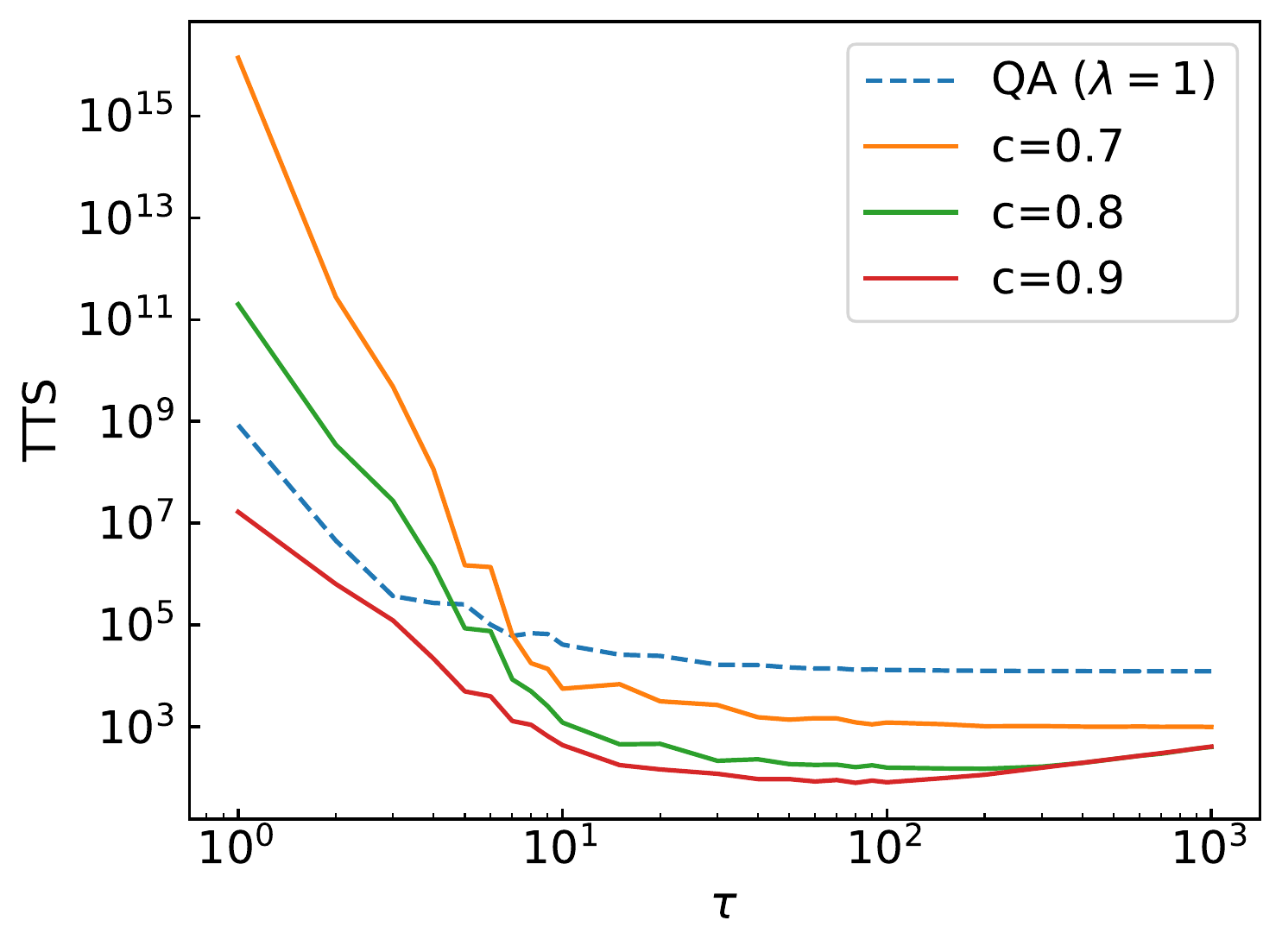}
  }
  \caption{TTS for $N=45$ and $p=3$ as a function of $\tau$ for (a) \(\Gamma=1\) and (b) \(\Gamma=2\). In the case of larger $\Gamma$, the TTS for ARA is shorter than that for QA. 
  Note the existence of minima for \(\Gamma=2\).
  }
  \label{fig:TTSN45eachc}
\end{figure}

Results for the TTS are depicted in Fig.~\ref{fig:TTSN45eachc} as a function of $\tau$ for $\Gamma=1$ and $\Gamma=2$, and fixed system size. For a weaker transverse field $\Gamma=1$, ARA does not necessarily perform better than conventional QA. For a larger $\Gamma=2$, ARA has a shorter TTS than QA for the parameter values we have tested.

\begin{figure}
  \centering
  \includegraphics[scale=0.5]{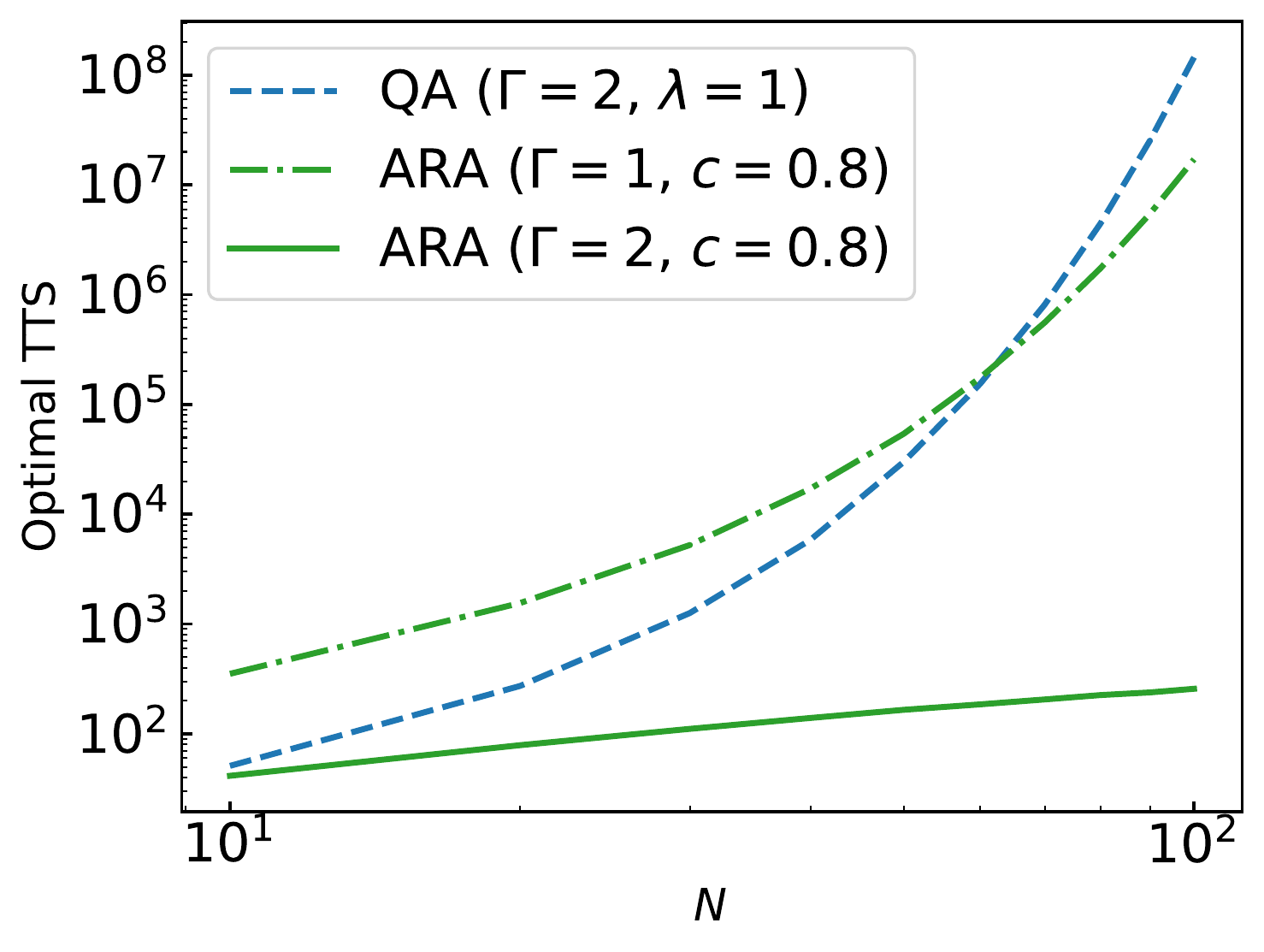}
  \caption{
  Size dependence of TTS of conventional QA (blue dashed line), ARA with \(\Gamma=1,~c=0.8\) (green dash-dot line), and \(\Gamma=2,~c=0.8\) (green solid line). As seen in the phase diagrams in Figs.~\ref{fig:phase-diag_G1} and~\ref{fig:phase-diag_G2}, ARA with \(\Gamma=1,~c=0.8\) encounters a first order transition whereas \(\Gamma=2,~c=0.8\) does not. This difference in statics is reflected in the dynamics as the exponential and polynomial dependence of the optimal TTS.
  }
  \label{fig:TTSscaling}
\end{figure}

Interestingly, minima exist in Fig.~\ref{fig:TTSN45eachc} for \(\Gamma=2\). This in fact holds for a broad range of system sizes, which allows us to extract the optimal annealing time $\tau$, $\tau_{\text{opt}} = \text{argmin}_{\tau} \text{TTS}(\tau, p_d)$ as a function of $N$, and hence the optimal TTS scaling with size~\cite{speedup,Albash:2017aa}. 
The optimal TTS, estimated from the data for $c=0.8$ as in Fig.~\ref{fig:TTSN45eachc}, is plotted in Fig.~\ref{fig:TTSscaling} as a function of system size. It is clearly seen that ARA with \(\Gamma=2,~c=0.8\) exhibits polynomial scaling, whereas ARA with \(\Gamma=1,~c=0.8\) and  conventional QA are exponential~
\footnote{Since this scaling is derived without the existence of an optimal annealing time, the true scaling can only be \emph{worse}, as shown in Ref.~\cite{Hen:2015rt}. Essentially, the reason is that the optimal TTS at small sizes $N$ is smaller than the TTS shown.}. 
This is consistent with the static phase diagram of Figs.~\ref{fig:phase-diag_G1} and~\ref{fig:phase-diag_G2}, where the path $s=\lambda$ does not cross a first order phase transition if \(\Gamma=2,~c=0.8\) in Fig.~\ref{fig:phase-diag_G2}, but conventional QA along $\lambda=1$ and ARA with \(\Gamma=1,~c=0.8\) in Fig.~\ref{fig:phase-diag_G1} do. This is quite non-trivial because the TTS is a purely dynamical measure for finite-size systems, whereas the static phase diagram represents the long-time and thermodynamic limits, $\tau\to\infty,~N\to\infty$.

\subsection{Classical spin vector dynamics}

It is instructive to compare the results of the  previous subsections with those of a counterpart classical model. In the spin-vector dynamics (SVD) model~\cite{Smolin, albash_reexamining_2015, Shin2014, muthukrishnan_tunneling_2016}, the system evolves according to the 
classical Hamilton dynamics under the semi-classical potential $V_{\rm SC}$ corresponding to the Hamiltonian of Eq.~(\ref{eq:ARA_H}),
%\DL{Were the angles $\phi_1$ and $\phi_2$ used anywhere in the simulations? Note that the $\phi$'s were introduced in \cite{albash_reexamining_2015} in deriving the SVD model from the path integral formulation (section 6), but they aren't necessary for our purposes and can be dropped as far as I can tell, and indeed were absent in the original SVD paper~\cite{Smolin} and its sequel~\cite{Shin2014}} \HN{Declared to be $\phi=0$},
\begin{equation}
 \begin{split}
   V_{\text{SC}} \equiv& \bra{\Omega(s)}\hat H(s)\ket{\Omega(s)}\\
    =&-sN\left(n_1\sin \theta_1 \cos \phi_1 + n_2\sin \theta_2 \cos \phi_2\right)^p\\
   &-(1-s)(1-\lambda)N(n_1\sin \theta_1 \cos \phi_1 - n_2 \sin \theta_2 \cos \phi_2)\\
 &-\Gamma (1-s)\lambda N(n_1\cos\theta_1 + n_2 \cos \theta_2),
\end{split}
 \label{ra:eq:semipote}
\end{equation}
where
\begin{align}
    n_1=\frac{Nc}{N},~n_2=\frac{N-Nc}{N},
\end{align}
and $\ket{\Omega(s)}$ is the spin-coherent state,
\begin{equation}
  \begin{split}
  &\ket{\Omega(s)} \\
  &=\bigotimes_{i=1}^{Nc} \qty[\cos\frac{\theta_1(s)}{2}\ket{-}_i + \sin\frac{\theta_1(s)}{2}e^{i\phi_1}\ket{+}_i] \\
  &\times \bigotimes_{i={Nc + 1}}^{N} \qty[\cos\frac{\theta_2(s)}{2}\ket{-}_i + \sin\frac{\theta_2(s)}{2}e^{i\phi_2}\ket{+}_i].
  \end{split}
\end{equation}
%The initial condition $\phi_1=\phi_2=0$ keeps these parameters $0$ throughout.
The initial conditions are $\phi_1=\phi_2=0$ and $\theta_1=-\theta_2=\pi/2$.
We used the equation of motion, which is derived from the path integral formulation and permutation symmetry \cite{muthukrishnan_tunneling_2016}.

\begin{figure}
 \centering
 \subfigure[\ \(\Gamma=1\)]{\includegraphics[scale=0.5]{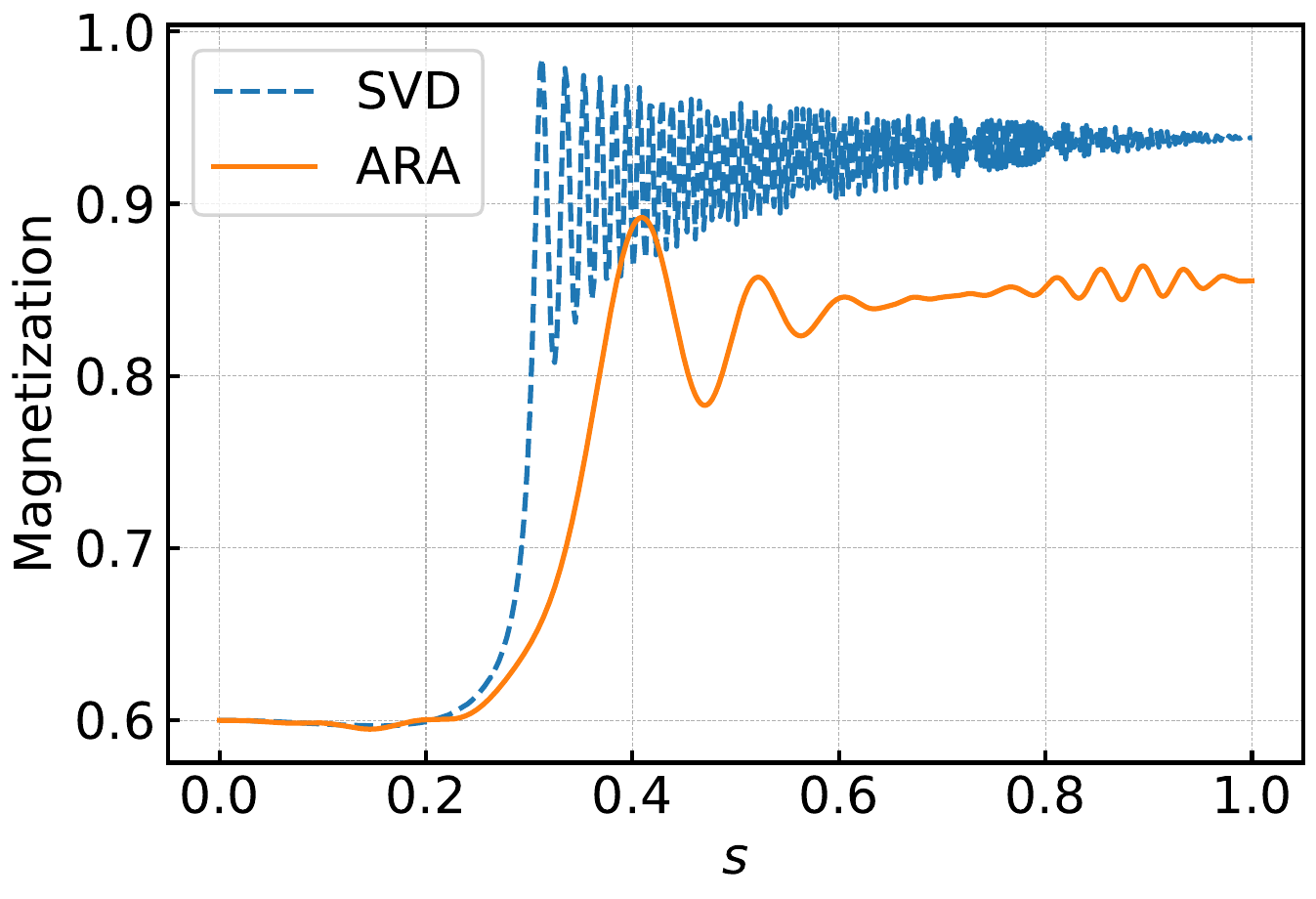}\label{fig:SVD-G1}}
 \subfigure[\ \(\Gamma=2\)]{\includegraphics[scale=0.5]{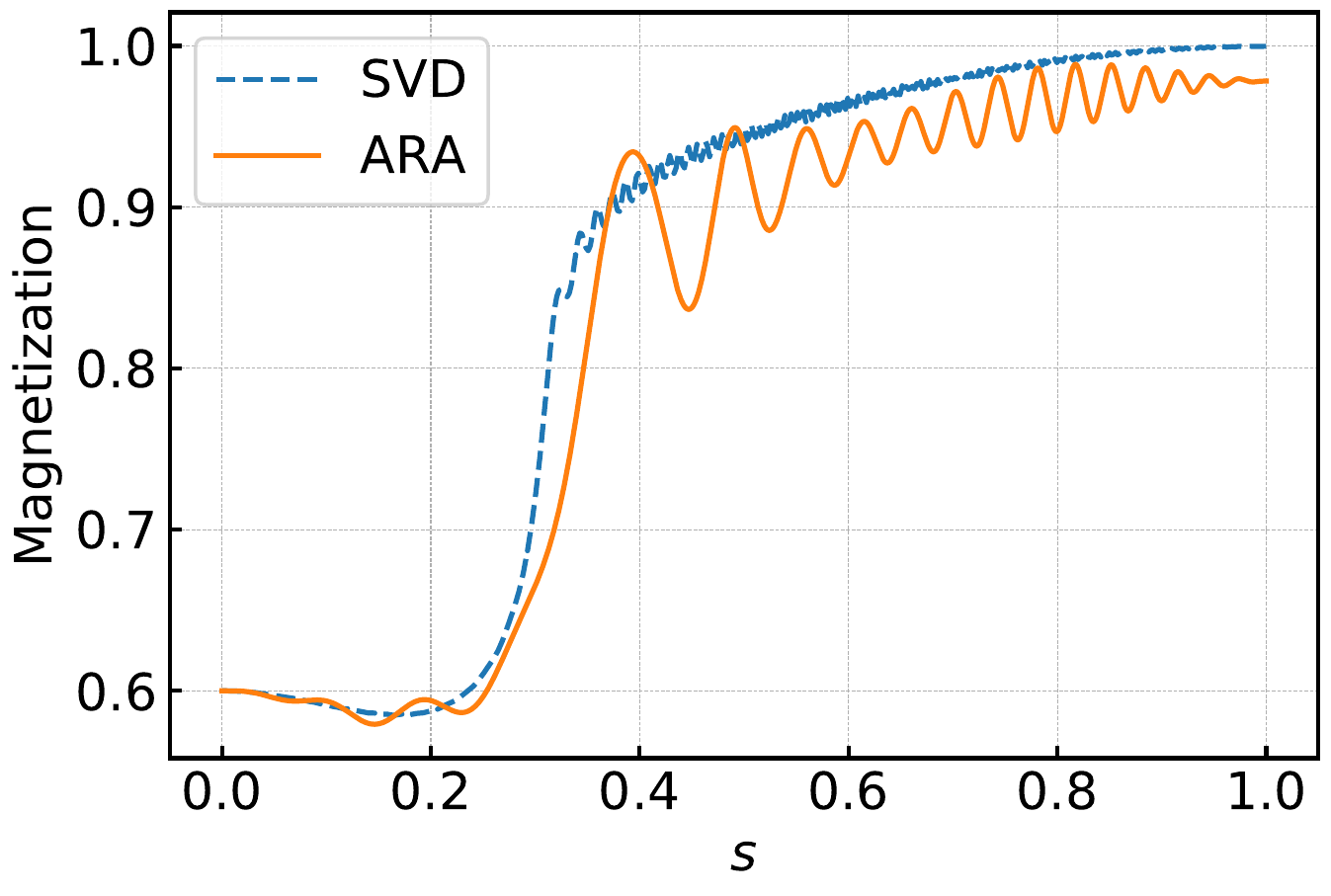}\label{fig:SVD-G2}}
 \subfigure[\ \(\Gamma=4\)]{\includegraphics[scale=0.5]{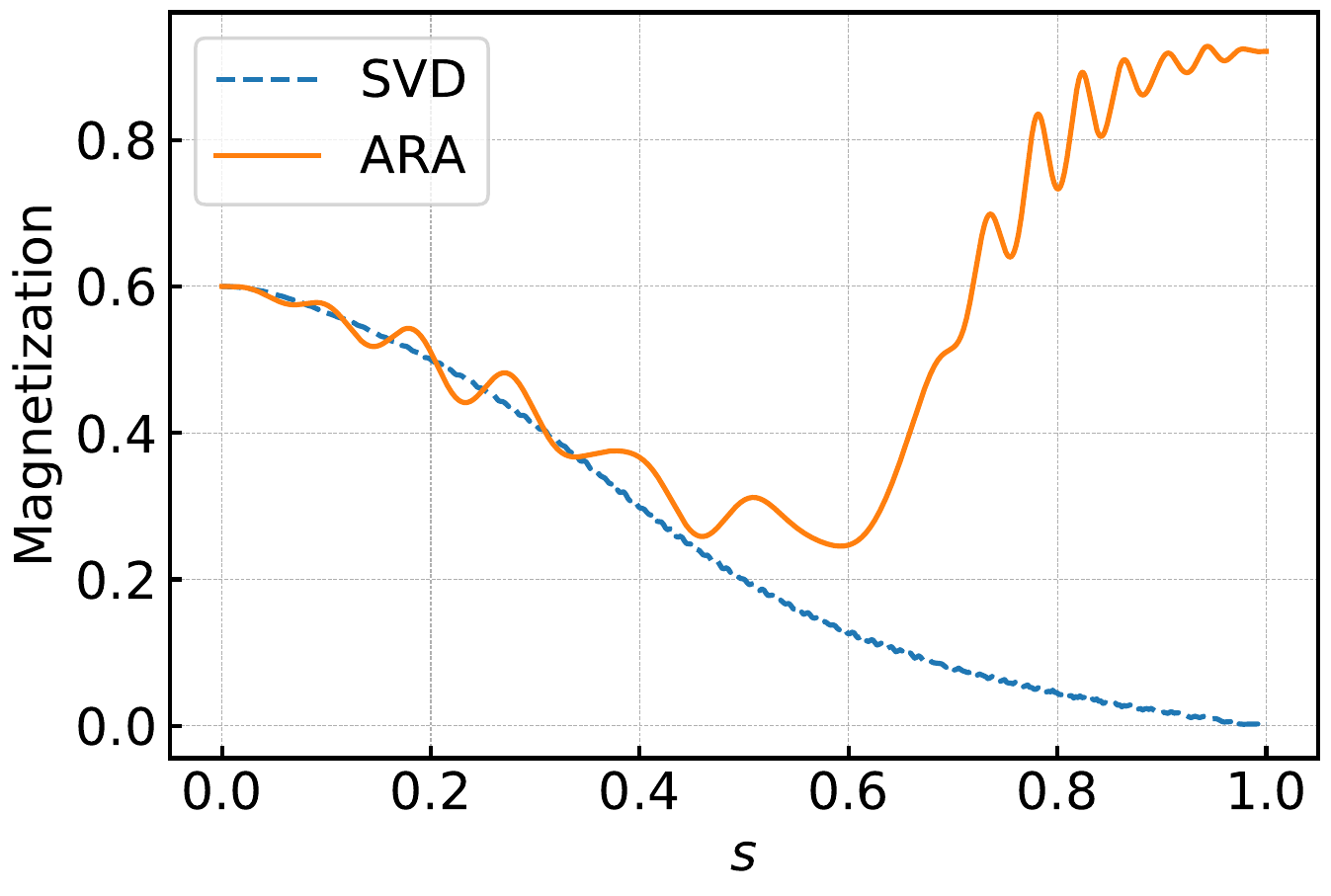}\label{fig:SVD-G4}}
\caption{Time evolution of the magnetization under ARA (orange) and SVD (blue, dashed) as a function of the normalized annealing time \(s\) for (a) $\Gamma=1$, (b) $\Gamma=2$, and (c) $\Gamma =4$ with $N=50, c=0.8$, $p=3$ and $\tau=40$.}
 \label{fig:SVD}
\end{figure}

The time evolution of the magnetization $\bra{\Omega (s)}{\frac{2}{N}(\hat S_1^z + \hat S_2^z)}\ket{\Omega(s)}$
under the SVD model is compared with its ARA quantum counterpart in Fig.~\ref{fig:SVD}. 
No essential differences are observed between classical and quantum dynamics in Figs.~\ref{fig:SVD-G1} and~\ref{fig:SVD-G2}.  However, in Fig.~\ref{fig:SVD-G4} the quantum dynamics succeeds in coming close to the right answer $m=1$ at the end, but the classical dynamics fails. 

\begin{figure}
	\centering
	\subfigure[\ $\Gamma=2,~s=\lambda=0.2$]{\includegraphics[scale=0.3]{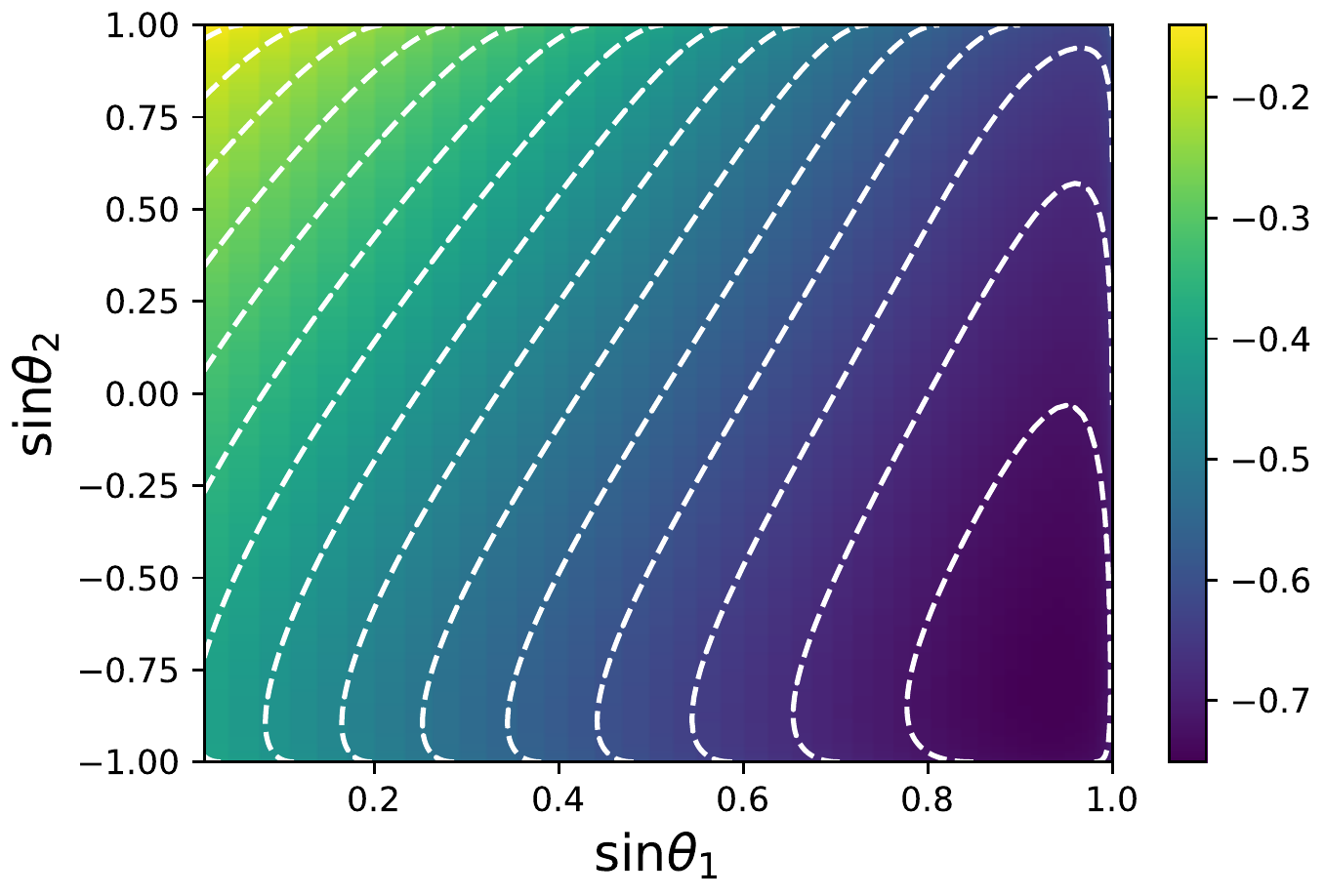}}
	\subfigure[\ $\Gamma=2,~s=\lambda=0.3$]{\includegraphics[scale=0.3]{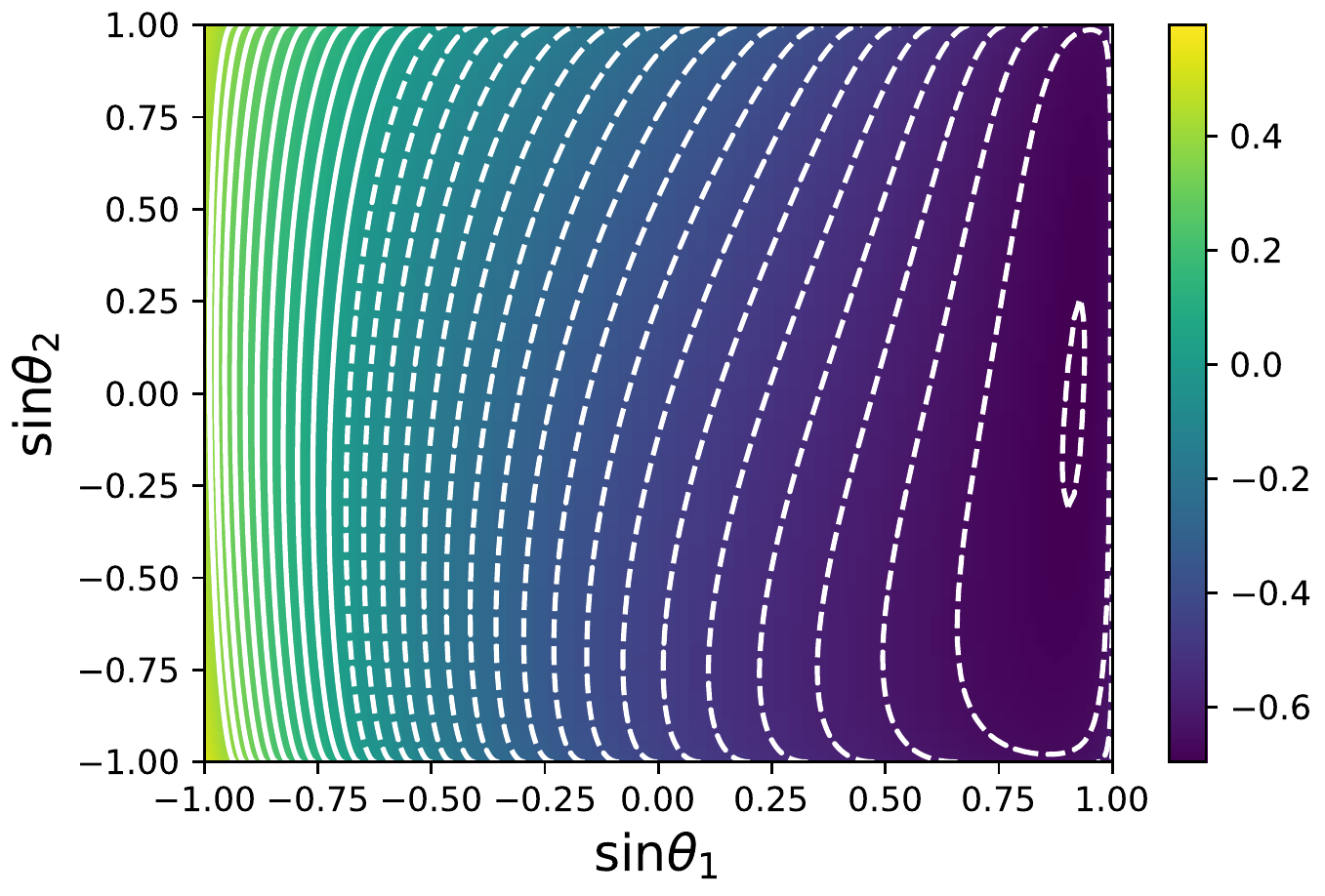}}
	\subfigure[\ $\Gamma=3.4,~s=\lambda=0.3$]{\includegraphics[scale=0.3]{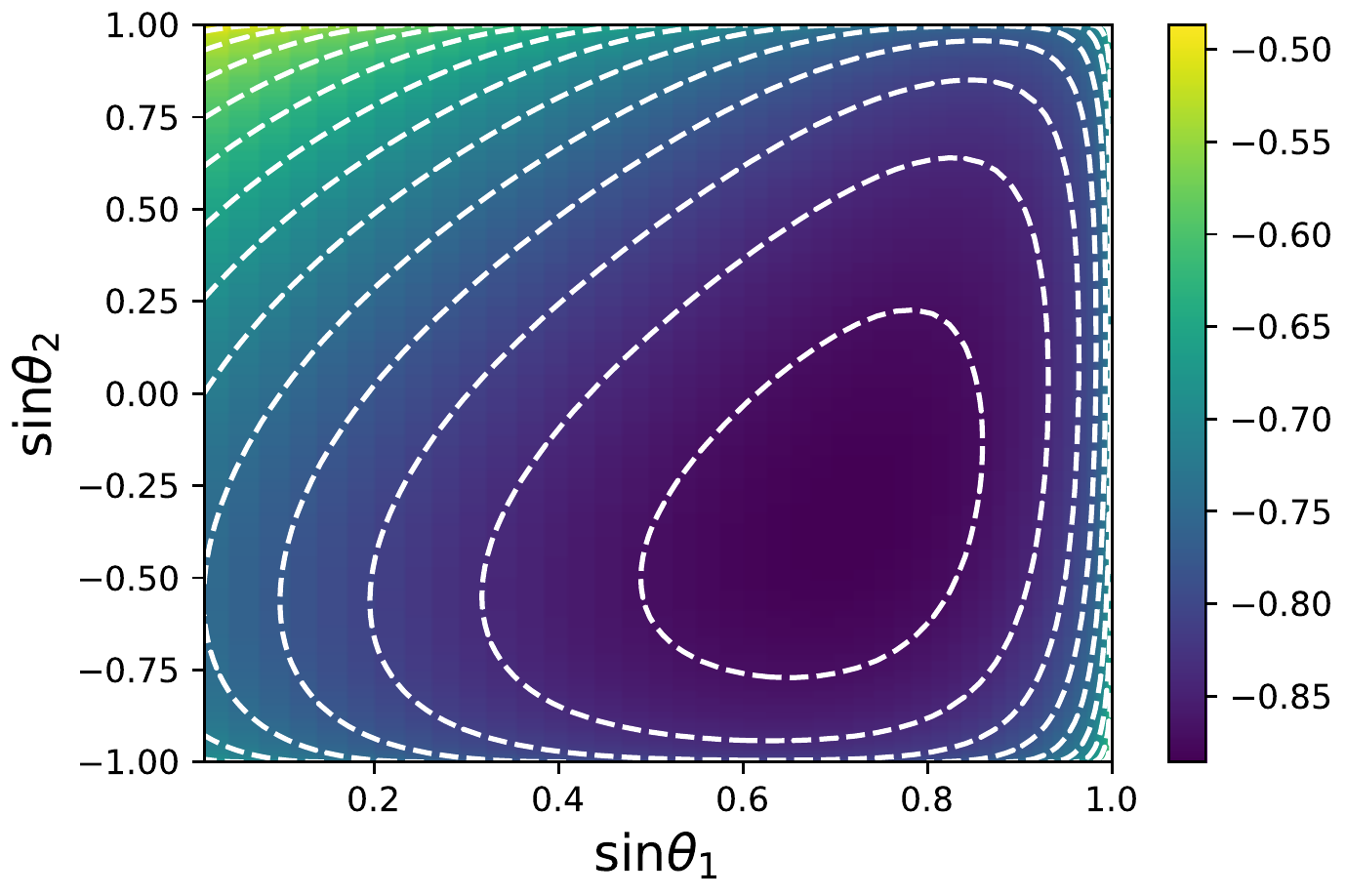}}
	\subfigure[\ $\Gamma=3.4,~s=\lambda=0.49$]{\includegraphics[scale=0.3]{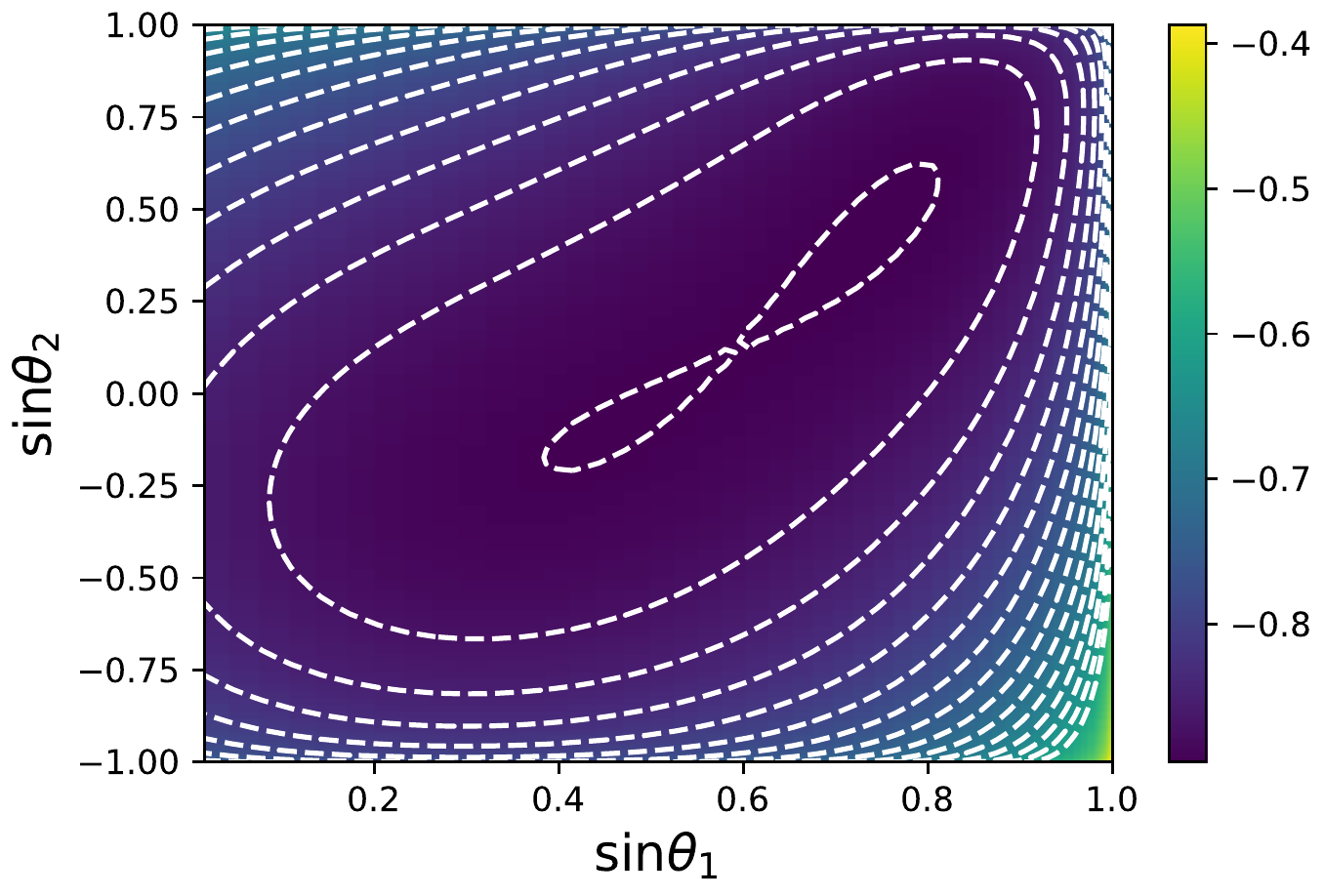}}
	\caption{Contour plot of the semi-classical potential $V_{\rm SC}$ as a function of the angles $\sin\theta_1$ and $\sin\theta_2$. Panels (a) and (b) for $\Gamma=2$ show that the minimum of the potential evolves continuously as a function of time $s(=\lambda)$ without encountering a phase transition in agreement with the phase diagram Fig.~\ref{fig:phase-diag_G2}. This leads to a large value of the magnetization at the end of the anneal as in Fig.~\ref{RA:fig:ARA_SVD}. By contrast, if $\Gamma$ is larger than a threshold value close to $3$ as in panels (c) and (d), the minimum at an earlier time $s=\lambda=0.3$ in (c) splits into two (almost) degenerate minima as in (d), and the system has to jump from a local minimum near the center of the figure to the global minimum at the right-top. This jump is possible by quantum tunneling since the barrier width and height are not very large at these parameter values, but is impossible for the classical SVD algorithm, resulting in the difference seen in Fig.~\ref{RA:fig:ARA_SVD}.
	}
	\label{fig:potential}
\end{figure}

\begin{figure}[htb]
  \centering
  \includegraphics[scale=0.5]{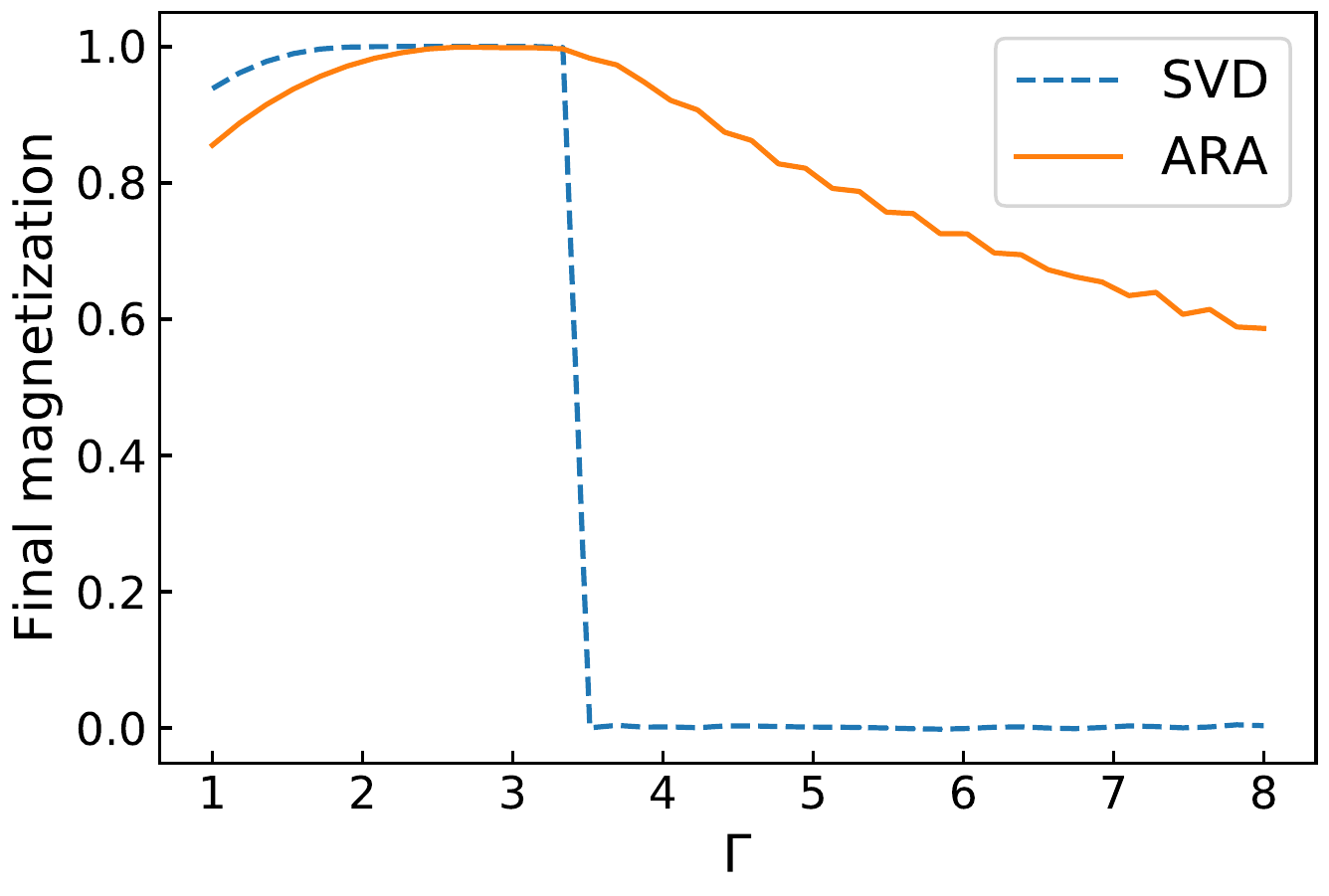}
  \caption{Final magnetization as a function of the strength of transverse field $\Gamma$ for $N=50$, $\tau=40$, $c=0.8$, and $p=3$.  The orange solid line represents the case of ARA and blue-dashed is for SVD.}
  \label{RA:fig:ARA_SVD}
\end{figure}

To explain this behavior, we note that for the parameter values of Fig.~\ref{fig:SVD-G4} [which lies between what is shown in Figs.~\ref{fig:phase-diag_G2} and~\ref{fig:phase-diag_G5}], the evolution path crosses the first order phase transition line near its termination in the middle of the $\lambda$-$s$ phase diagram, where the energy barrier across the first order transition is thin and low. Quantum dynamics apparently tunnels through the barrier but the classical SVD algorithm gets stuck because there is no classical mechanism to go through or over the barrier, however thin or low the barrier is.
Figure~\ref{fig:potential} supports this viewpoint, by illustrating the appearance of an energy barrier in the semiclassical potential.  Figure~\ref{RA:fig:ARA_SVD} conspicuously demonstrates the marked difference between classical and quantum dynamics, which is not clear in the static phase diagram. It shows the final value of the magnetization as a function of $\Gamma$. Quantum dynamics exhibits a gradual deterioration as $\Gamma$ increases beyond $3.4$, corresponding to a lower tunneling rate, whereas the classical SVD suddenly fails beyond a threshold close to $\Gamma=3.4$, where the energy barrier seen in Fig.~\ref{fig:potential} appears. %
\subsection{Classical spin vector Monte Carlo}

\begin{figure}
 \centering
 \subfigure[\(\beta=10\)]{\includegraphics[scale=0.47]{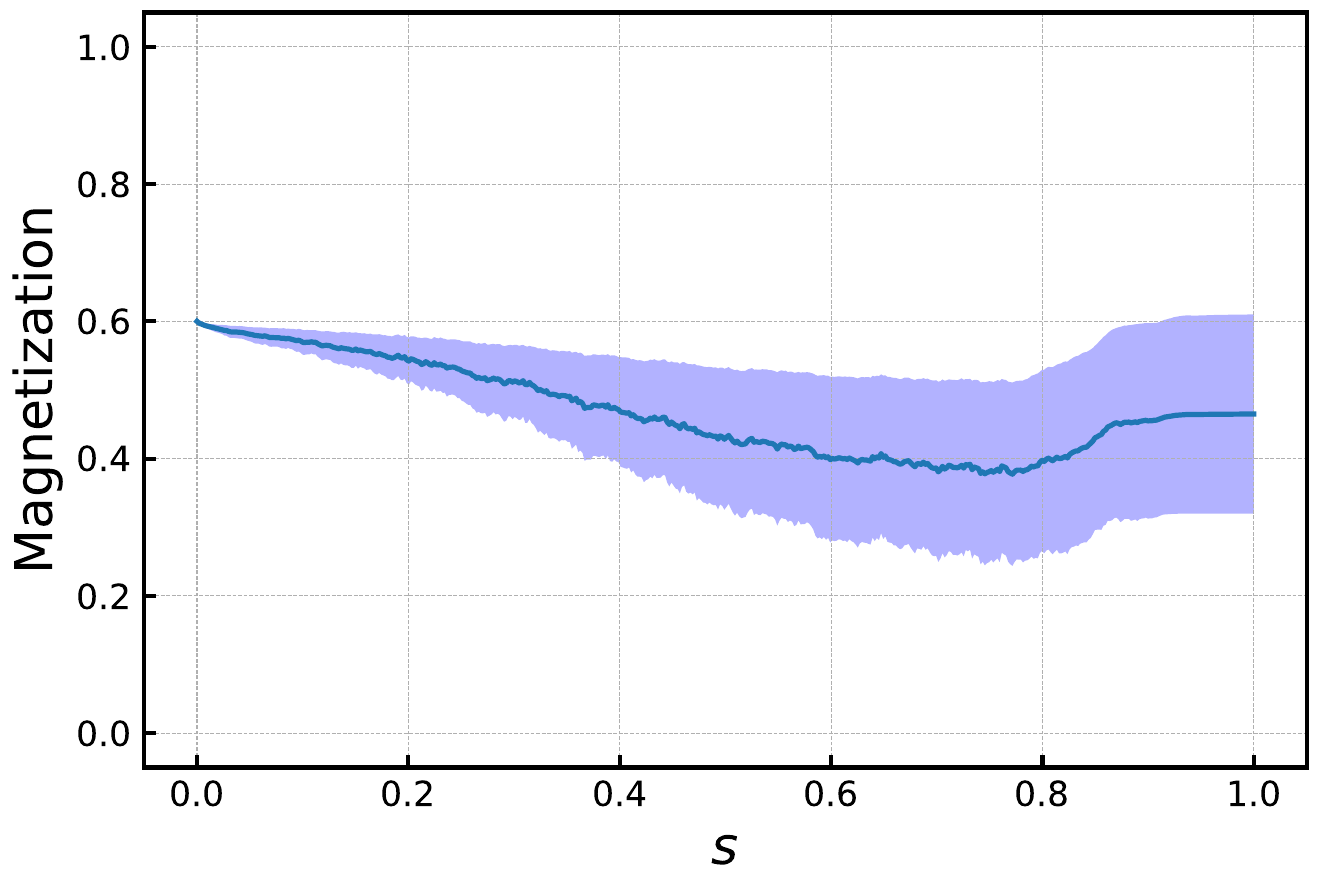}\label{fig:SVMCbeta10}}
 \subfigure[\(\beta=5\)]{\includegraphics[scale=0.47]{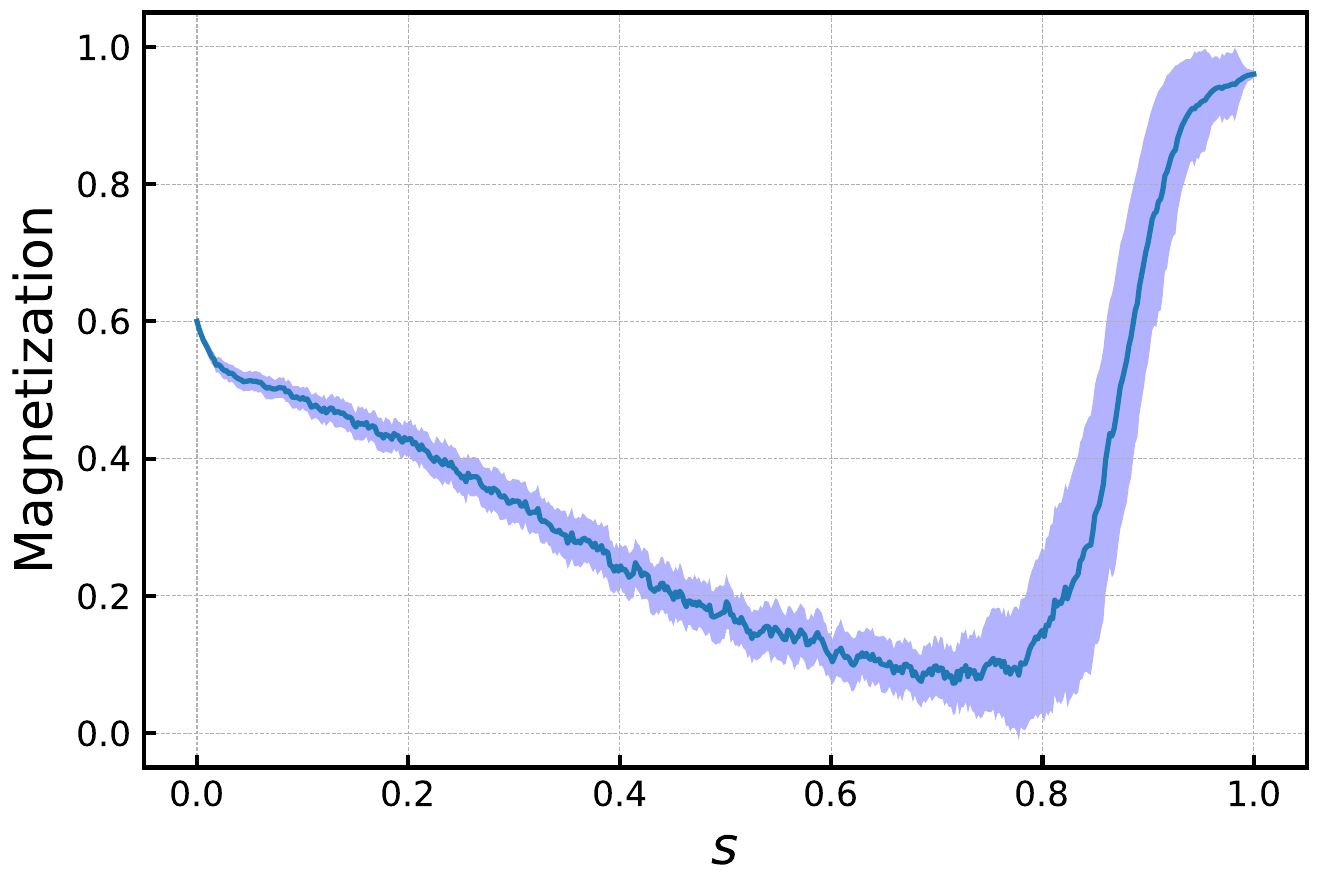}\label{fig:SVMCbeta5}}
 \subfigure[\(\beta=1\)]{\includegraphics[scale=0.47]{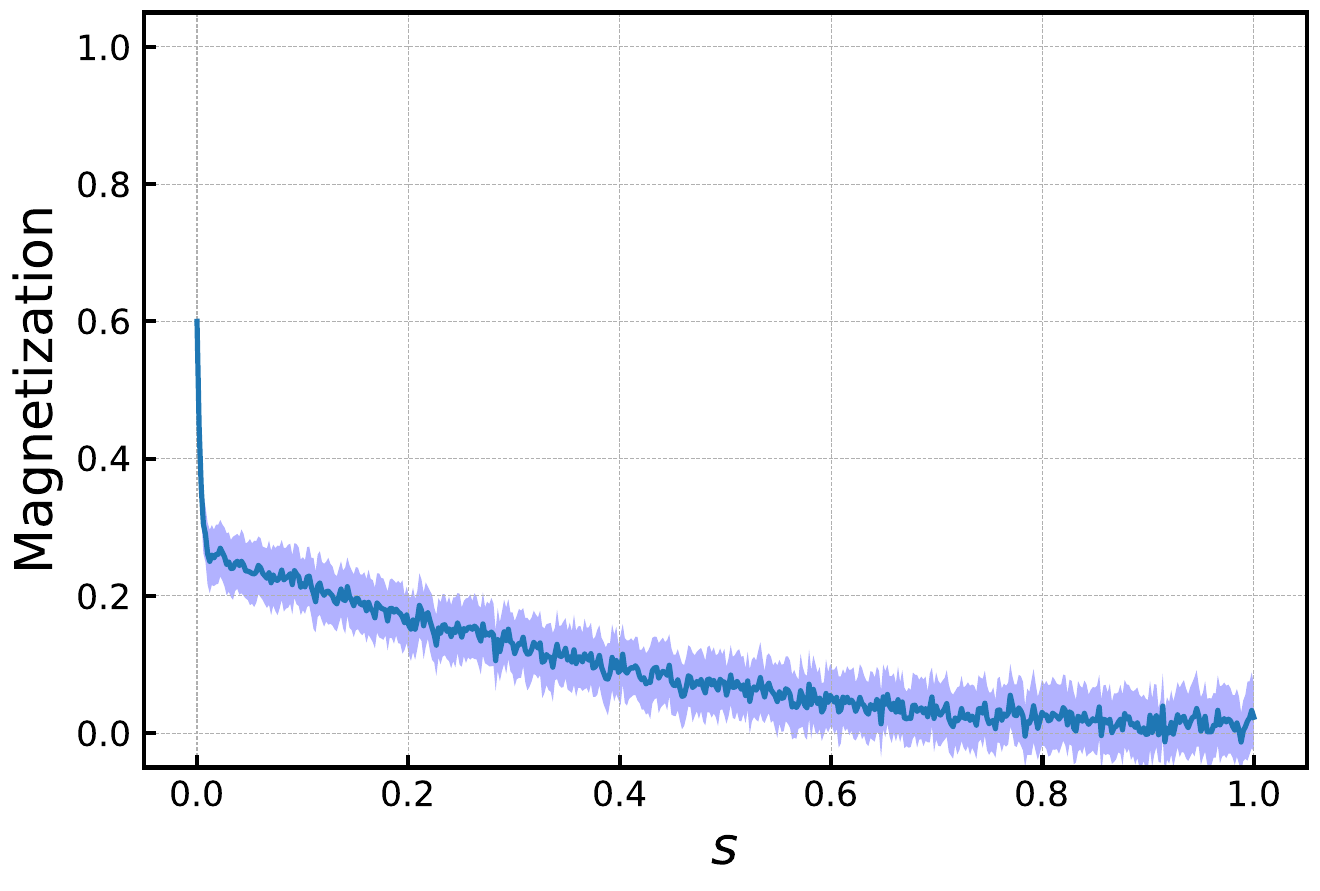}\label{fig:SVMCbeta1}}
\caption{Time evolution of the magnetization under spin vector Monte Carlo at finite temperature $\beta^{-1}$ with $N=50, c=0.8$,  $\Gamma=4$, and $500$ sweeps (a sweep is a complete Metropolis update of all the spins at a given $s$). The width of the light blue hue is the standard deviation calculated from 100 runs.  (a) Since $\beta = 10$ is too large (temperature too low) for the system to escape the basin of the initial state, the magnetization just wiggles around the initial value. (b) At a higher temperature $\beta = 5$, the state can hop over the energy barrier to reach the correct ferromagnetic state in the end. (c) If the temperature is too high $\beta=1$, the state becomes completely random.}
 \label{fig:SVMC}
\end{figure}

Since the energy barrier is thin and low for Fig.~\ref{fig:SVD}(c), the classical system can hop over the barrier if we introduce thermal fluctuations.  Figure~\ref{fig:SVMC} shows the results of spin vector Monte Carlo~\cite{Shin2014}, a finite-temperature version of SVD, in which the angle variables in the semiclassical potential are updated stochastically according to the usual Monte Carlo rule.
As seen in this figure, the system succeeds in reaching a good answer, $m$ close to 1, if the temperature is chosen appropriately, but if the temperature is too high, the final value is far from 1. Clearly, the temperature needs to be tuned in order to reach the correct solution. This is simple to do in the present case, but in general one does not know the right value in advance.  This is in contrast with the quantum dynamics, where there is no such adjustable parameter.

\section{Iterated Reverse annealing}
\label{Sec:iRA}

\begin{figure}[h]
  \centering
   \includegraphics[scale=0.55]{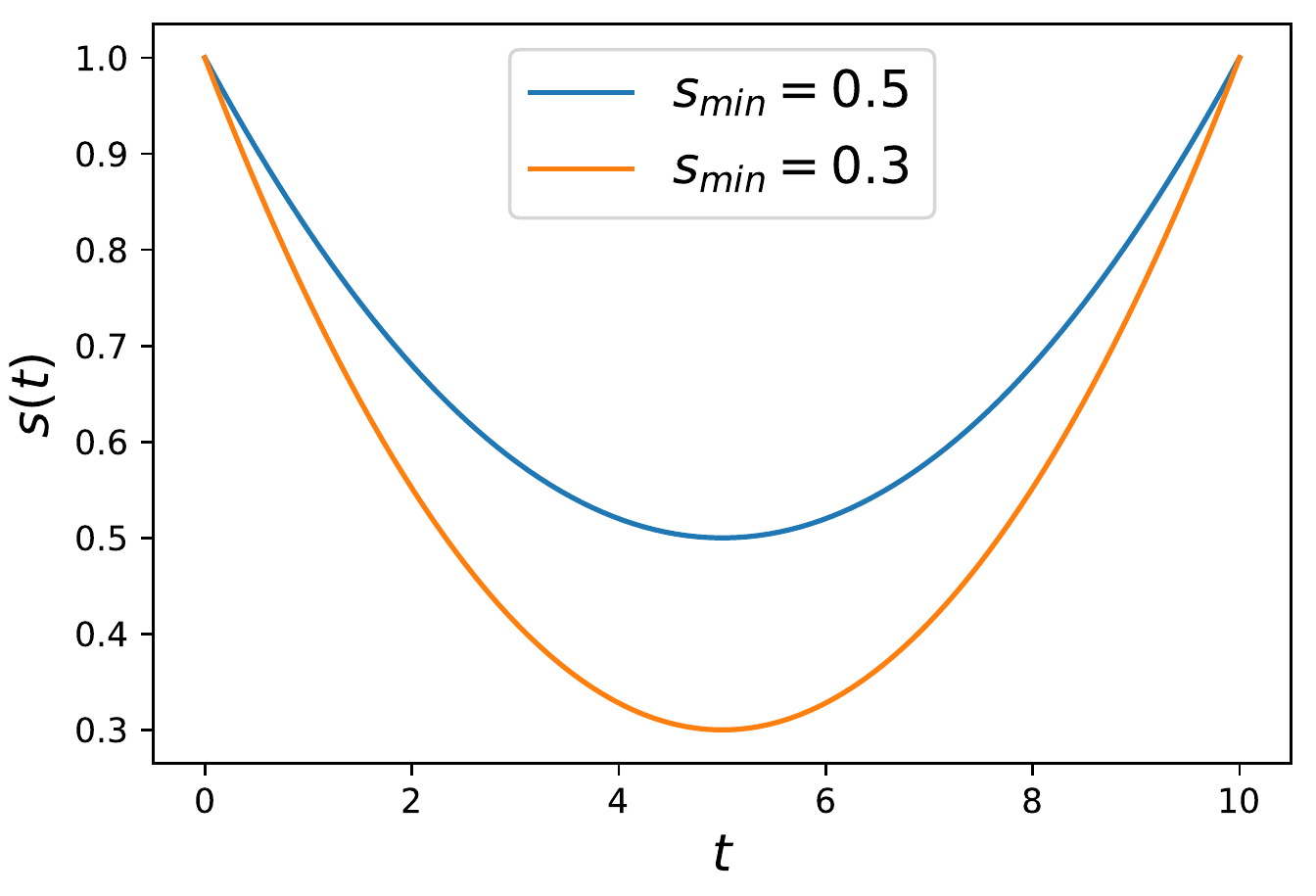}
 \caption{Annealing schedule of a single cycle of IRA for $\tau=10$. The schedule in IRA is non-monotonic as a function of time, starting and ending at $s=1$.}
    \label{fig:iRA-schedule}
\end{figure}

In iterated reverse annealing (IRA), one measures the state in the computational (classical) basis after a single cycle of RA and starts the next cycle from the classical state thus obtained. 
%More formally, a single cycle of this algorithm is
%\begin{equation}
%\rho_{\text{init}} \xrightarrow{{\rm RA}}
%\hat U \rho_{\text{init}} \hat U^\dagger  \xrightarrow{{\rm meas.}}
%\rho_{\text{next}}=\sum_i \hat P_i(\hat U \rho_{\text{init}} \hat U^\dagger)\hat P_i
%\label{eq:iRA_alg}
%\end{equation}
%where \(\hat P_i\) is the projection onto a classical state $\ket{i}$, where $i$ is the decimal representation of a bit string representing the states of all the qubits after measurement, and $p(i) = \Tr[\hat P_i(\hat U \rho_{\text{init}} \hat U^\dagger)\hat P_i]$ is the probability that this state was obtained. 
 Let \(\hat U\) denote the time evolution operator under the Hamiltonian of reverse annealing without the initialization term present in ARA,
\begin{equation}
 \hat H(t) = A(t)\hat H_0 + B(t)\hat V_{\text{TF}}
 \label{eq:iRA_H}
\end{equation}
where \(A(t) = s(t), B(t)=1-s(t)\) with an appropriate choice of the function $s(t)$. The key difference from standard QA is that now $s(t)$ is non-monotonic and has a minimum, $s_{\min}$, as shown in Fig.~\ref{fig:iRA-schedule}.
The value of this minimum relative to the critical value $s_c$, plays an important role, since it determines whether or not the system crosses a phase transition: it does when $ s_{\min} < s_c$. 

Time changes from $t=0$ to \(t=\tau\), where \(\tau\) is the annealing time of a single cycle, and $s(0) = s(\tau)=1$. The total annealing time with \(r\) cycles is \(r\tau\).  The initial state is classical, i.e., a computational basis state $\ket{i}$ (where $i\in[0,\dots,2^N-1]$ is the state index) and the probability of the final state being $\ket{j}$ after a cycle is $p_{ji}(\tau) = \sum_{k\in S(e_j)} |\bra{k}\hat{U}(\tau)|i\rangle |^2$, where $S(e_j)$ is the index set of the degenerate states with energy $e_j$. This process is realized in the current version of the D-Wave 2000Q device, except that $\hat{U}$ is replaced by open quantum system evolution, which means that thermal effects play a role as well during each cycle in the real device.

IRA is an algorithm that repeats the cycle of RA and measurement. Therefore, IRA can be considered as a series of classical Markov transitions, and a single cycle is a stochastic transition to the final classical state from the initial classical state. Using the overlap $p_{ji}(\tau)$ above, the classical transition probability matrix $P$ has elements $P_{ji} = p_{ji}(\tau)$. Let $\pi$ denote a classical probability vector whose $i$th component $\pi_i$ denotes the probability of obtaining the $i$th computational basis state. After $r$ cycles, the probability becomes
\begin{equation}
\pi_j^{(r)} = \sum_i (P^r)_{ji} \pi_i^{(0)} ,
\end{equation}
where $\pi_i^{(0)}$ is the given initial state. The probability of obtaining the ground state is $\pi_0^{(r)}$. 

A sufficient condition for IRA to succeed is that the ground state probability increases after a single cycle. Correspondingly, excitations should be suppressed in a single cycle. Unfortunately, this condition does not seem to be satisfied for the $p$-spin model as shown below, at least for the specific schedule choice we tested.

\begin{figure}[t]
 \centering
 \subfigure[\ Distribution of the final magnetization for $s_{\min}=0.5$.]{
   \includegraphics[scale=0.5]{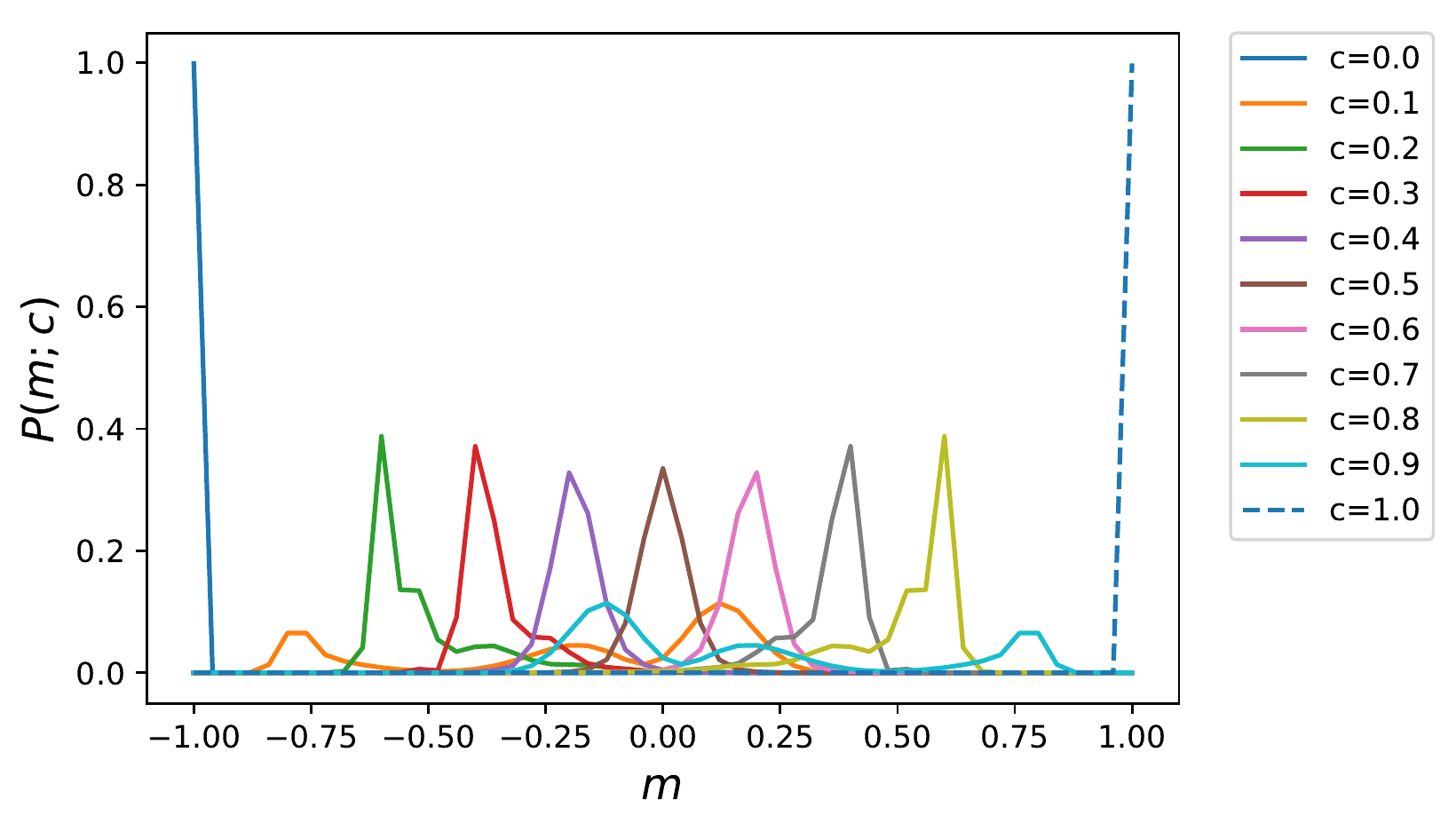}
   \label{ra:fig:mag_dist_05}
 }
 \centering
 \subfigure[\ Distribution of the final magnetization for $s_{\min}=0.3$.]{
   \includegraphics[scale=0.5]{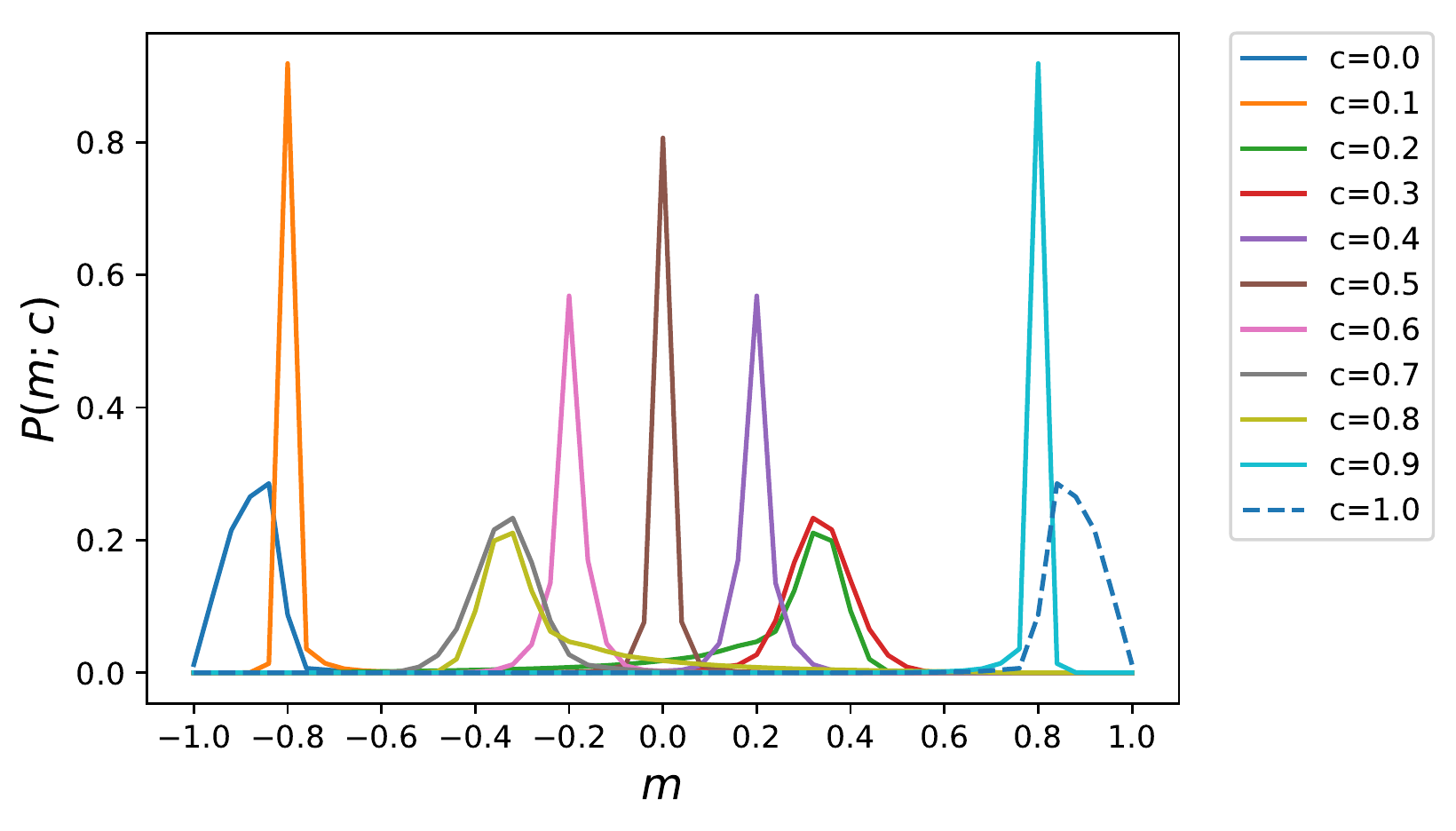}
   \label{ra:fig:mag_dist_03}
 }
 \caption{
 Distribution of magnetization of the final state with various initial conditions (indicated by the values of $c$) with two different minimum values of $s$ (above and below the critical $s$ value), and $N=50$ and $\tau=10$.
 }
 \label{fig:iRA}
\end{figure}

We choose a quadratic function of $s(t)$ with a minimum \(s_{\min}\) as in Fig.~\ref{fig:iRA-schedule}.
Figures~\ref{ra:fig:mag_dist_05} and~\ref{ra:fig:mag_dist_03} show the distribution of the magnetization of the final states after a single cycle, starting from various values of $c$ with $N=50$, $\tau=10$ and $p=3$. In Fig.~\ref{ra:fig:mag_dist_05}, since \(s_{\min}=0.5\) is above the critical value $s_c$,% 
\footnote{
See Fig.~3 (left panel) of Ref.~\cite{seki_quantum_2012}.}
and hence the system does not undergo a phase transition, the magnetization stays close to the initial value (given by $c$), though it is shifted somewhat towards $m=0$. If \(s_{\min}\) is below the transition point as in Fig.~\ref{ra:fig:mag_dist_03}, the system does cross the phase transition and, as expected, the final magnetization deviates more from the initial value. Specifically, the final magnetization moves toward $m=0$, especially for $c\in [0.3,0.7]$. 
%However, if the initial state is the ground state, the maximum excited state, and $c = 0.9, 0.1$, it seems to stay in the initial state.
However, if the initial state is the ground state, the system seems to stay near the initial state.
% It is observed, however, that the distribution shifts away from the solution $m=1$. \DL{How generic is this if the values of $N$ and $\tau$ are changed?} \HN{Data appended as additional figures.}

\begin{figure}[t]
\centering
  \subfigure[\ $c=1$]{
    {\includegraphics[scale=0.55]{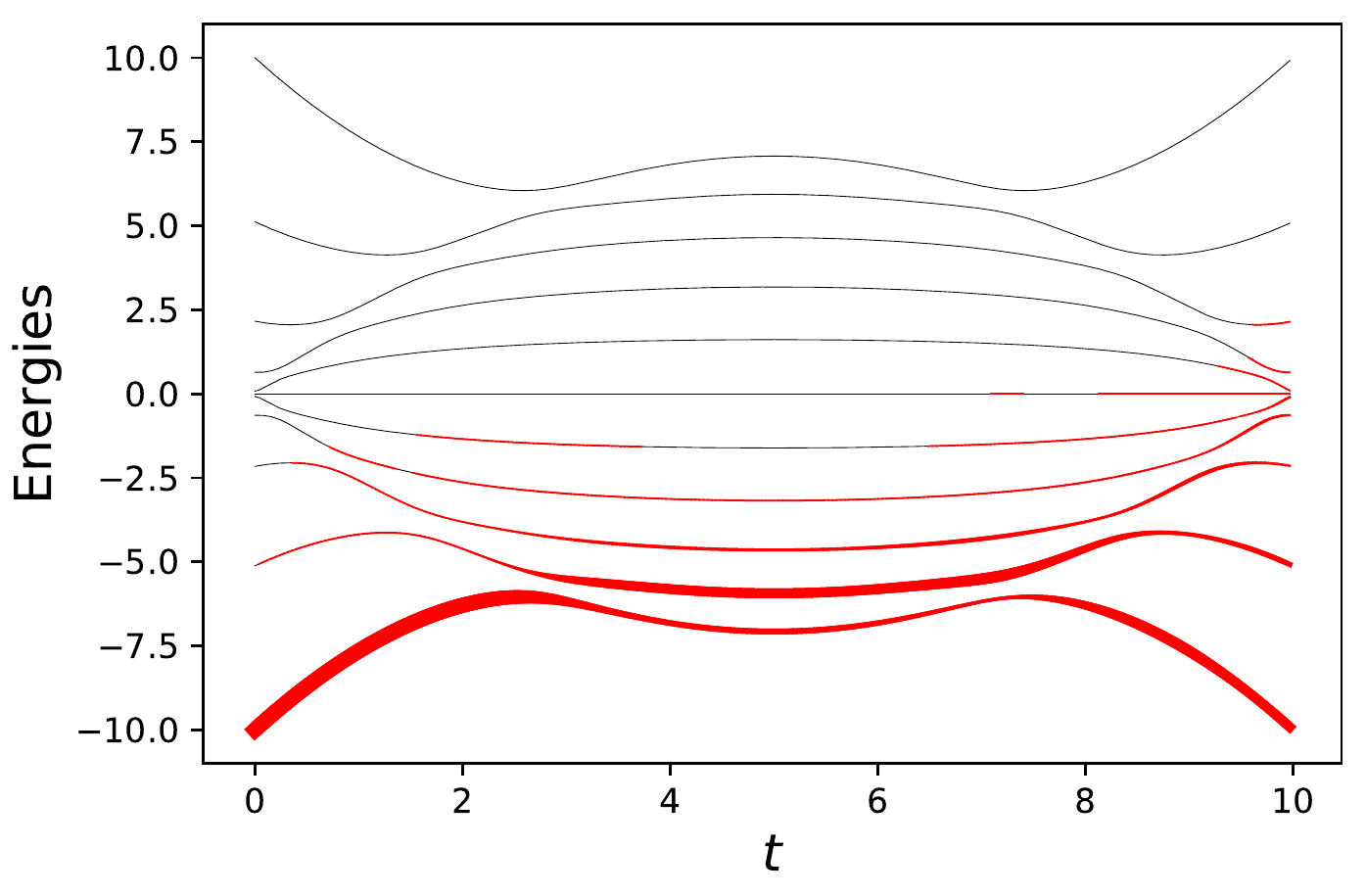}\label{fig:iRA_spect_a}}
  }
  \centering
  \subfigure[\ $c=0.8$]{
    {\includegraphics[scale=0.55]{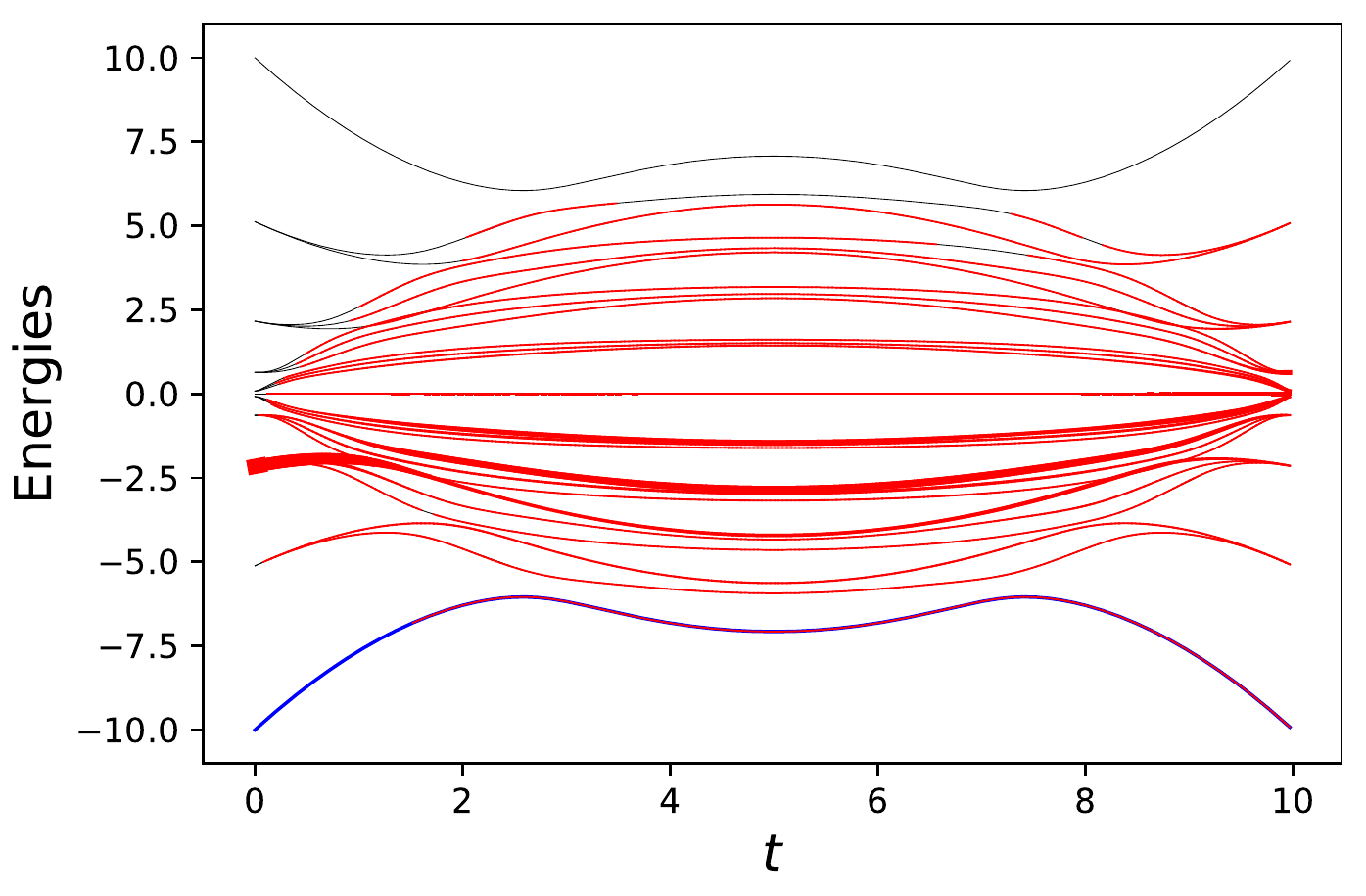}\label{fig:iRA_spect_b}}
  }
  \caption{Energy spectrum of IRA during a single cycle and occupation probability (represented by the red  thickness) for $N=10$, $p=3$, $\tau=10$ and $s_{\min}=0.3$.
The blue solid line represents the ground state and the black thin lines represent the excited states. In (a) the system starts in the ground state ($c=1$) and in (b) it starts in an excited state ($c=0.8$).
    }
  \label{fig:iRA_spect}
\end{figure}

To better understand what happens in Fig.~\ref{fig:iRA}, we have calculated the energy spectrum and occupation probability.
Results are shown in Fig.~\ref{fig:iRA_spect} for $s_{\min}=0.3$. 

Because of the symmetry of the annealing schedule of IRA with respect to $s=\frac{1}{2}$, avoided crossings in the first half appear also in the second half.
Therefore, when annealing starts from the ground state as shown in Fig. \ref{fig:iRA_spect_a}, the state becomes excited once and then returns to the ground state at the second avoided level crossing.
It can also happen that due to the large number of avoided crossings in the first and second halves, an excited state can be further excited to higher energy levels, as in Fig.~\ref{fig:iRA_spect_b}. The situation is far from simple, and the general tendency is that the distribution among states becomes broader as cycles are iterated except for the limiting case of adiabatic annealing, where the total evolution time $\tau$ is large on the scale set by the inverse of the minimum gap between the ground and first excited state.

\begin{figure}[htb]
\centering
  \subfigure[\ \(s_{\min}=0.5, \tau =30, r=1\)]{
    \includegraphics[scale=0.4]{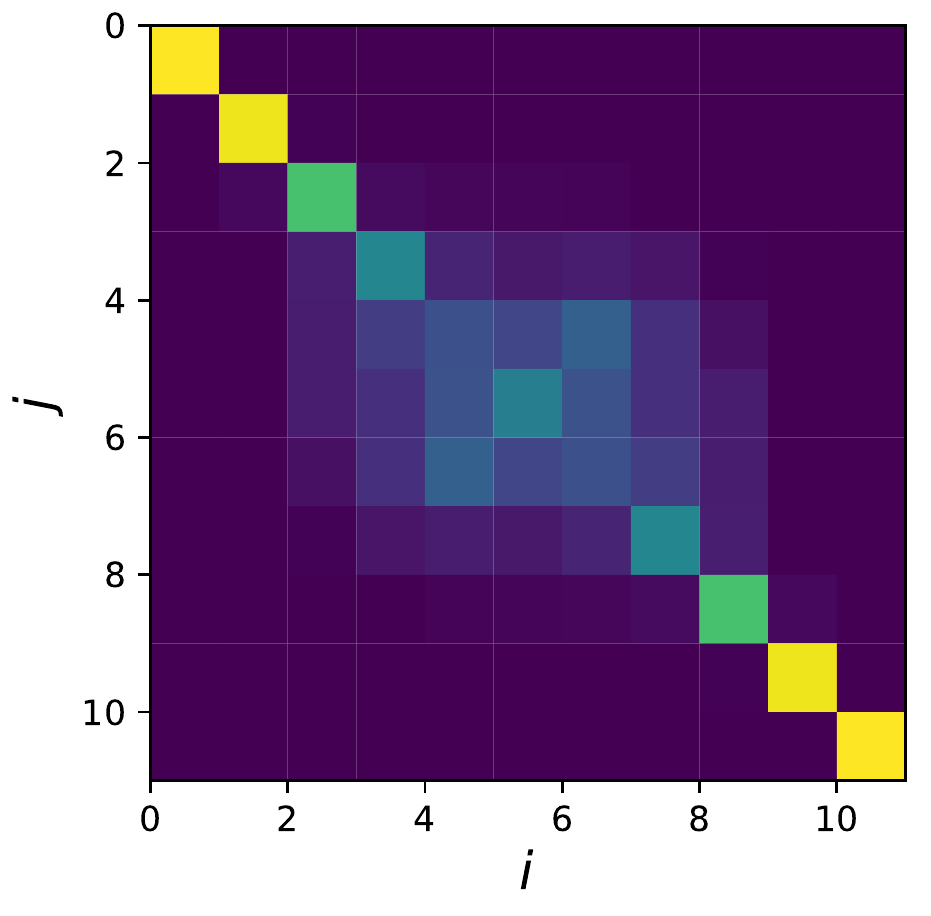}
    \label{fig:IRAtau30r1}
  }
   \subfigure[\ \(s_{\min}=0.5, \tau =30, r=5\)]{
    \includegraphics[scale=0.4]{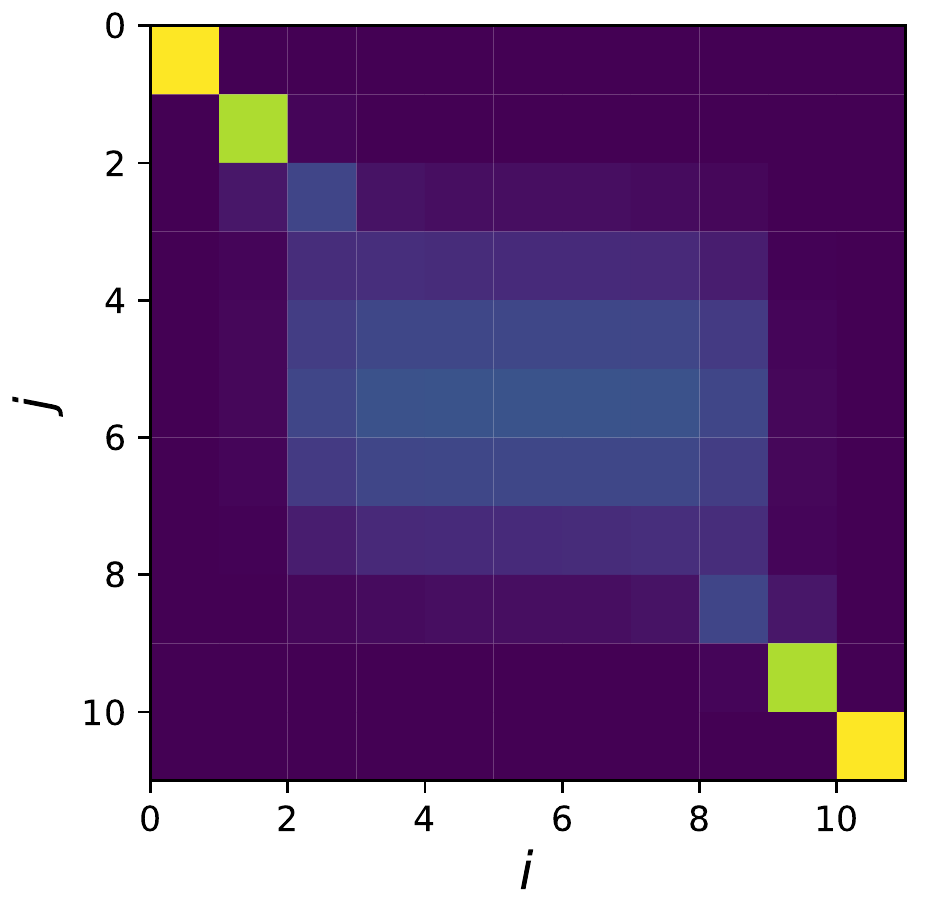}
    \label{fig:smin5IRAtau30r5}
  }
    \subfigure[\ \(s_{\min}=0.5, \tau =8, r=1\)]{
    \includegraphics[scale=0.4]{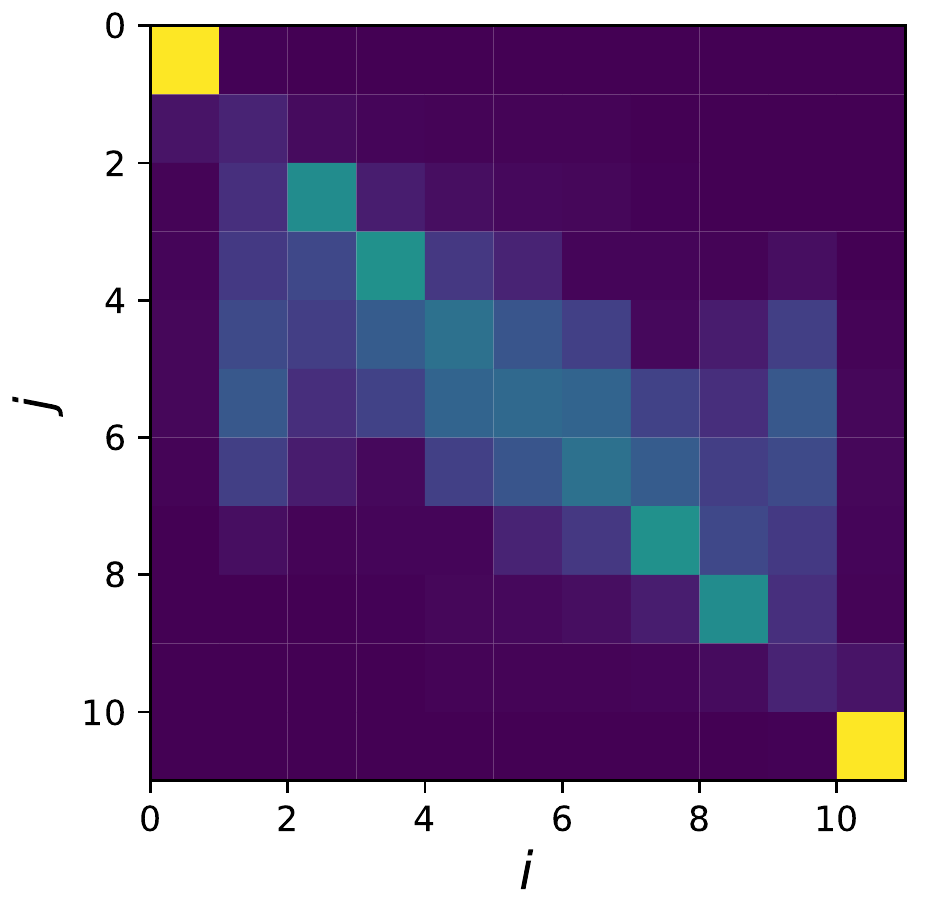}
    \label{fig:IRAtau8r1}
  }
   \subfigure[\ \(s_{\min}=0.5, \tau =8, r=3\)]{
    \includegraphics[scale=0.4]{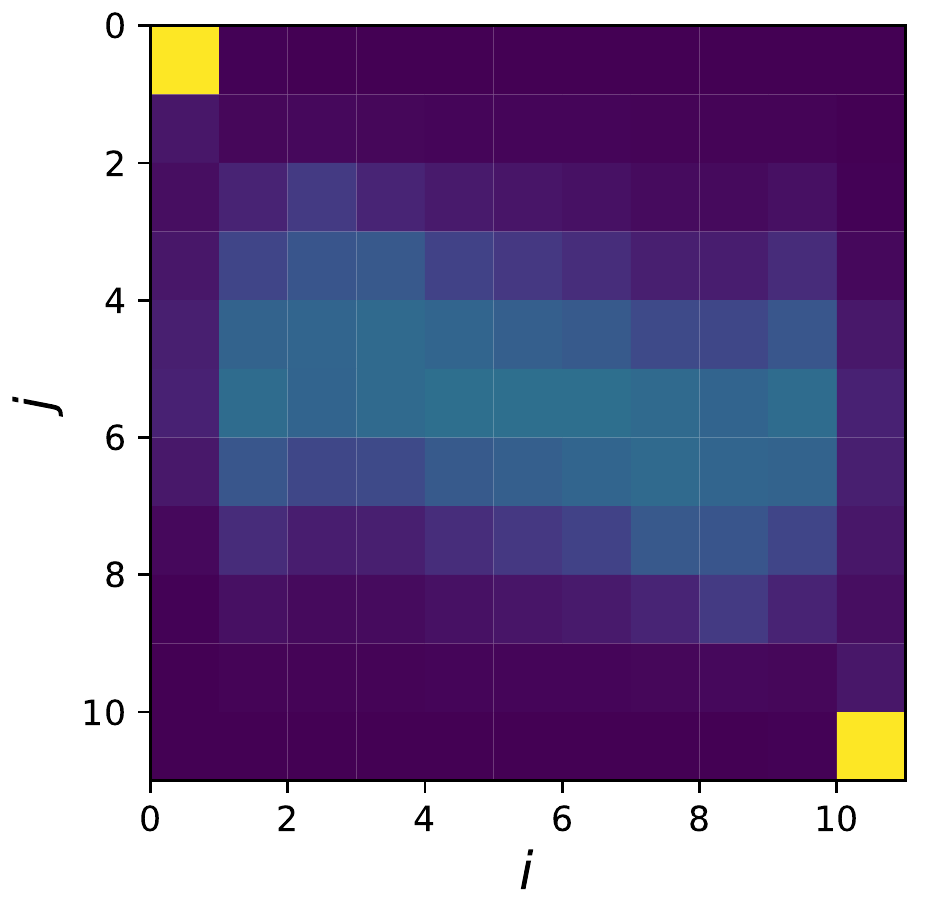}
    \label{fig:smin5IRAtau8r5}
  }
 
  \caption{Transition probabilities $(P^r)_{ji}$ from state $i$ (horizontal axis) to state $j$ (vertical axis) with $i=0$ the ground state, $i=1$ the first excited state, etc., for $p=3, ~N=10, ~s_{\min}=0.5$. Panels (b) and (d) are for the results after $r$ repetitions.
  In this case of $s _{\min} = 0.5$, the phase transition is not crossed, so the probability of staying in the ground state is high as seen in the left-top element in each panel. The general tendency is clear that transitions to the ground state, shown in the top row, are small except for the top-left corner which represents the probability to stay in the initially-given ground state.
  Panel (a) has a rather large $\tau$ and the system stays close to adiabatic, which is reflected in the high probabilities along the diagonal.
  In panel (c), with a smaller $\tau$, some diabatic transitions take place as seen by the increased brightness of the off-diagonal elements.
  Since $s_{\min}=0.5 > s_c$, the energy gap $\Delta_{01}$  between the ground state and the first excited state remains relatively large, but transitions do occur between excited states because the energy gaps between higher energy states are smaller than $\Delta_{01}$.
  }
    \label{fig:IRAsmin05P}
\end{figure}

\begin{figure}[htb]
\centering
  \subfigure[\ \(s_{\min}=0.3, \tau =30, r=1\)]{
    \includegraphics[scale=0.4]{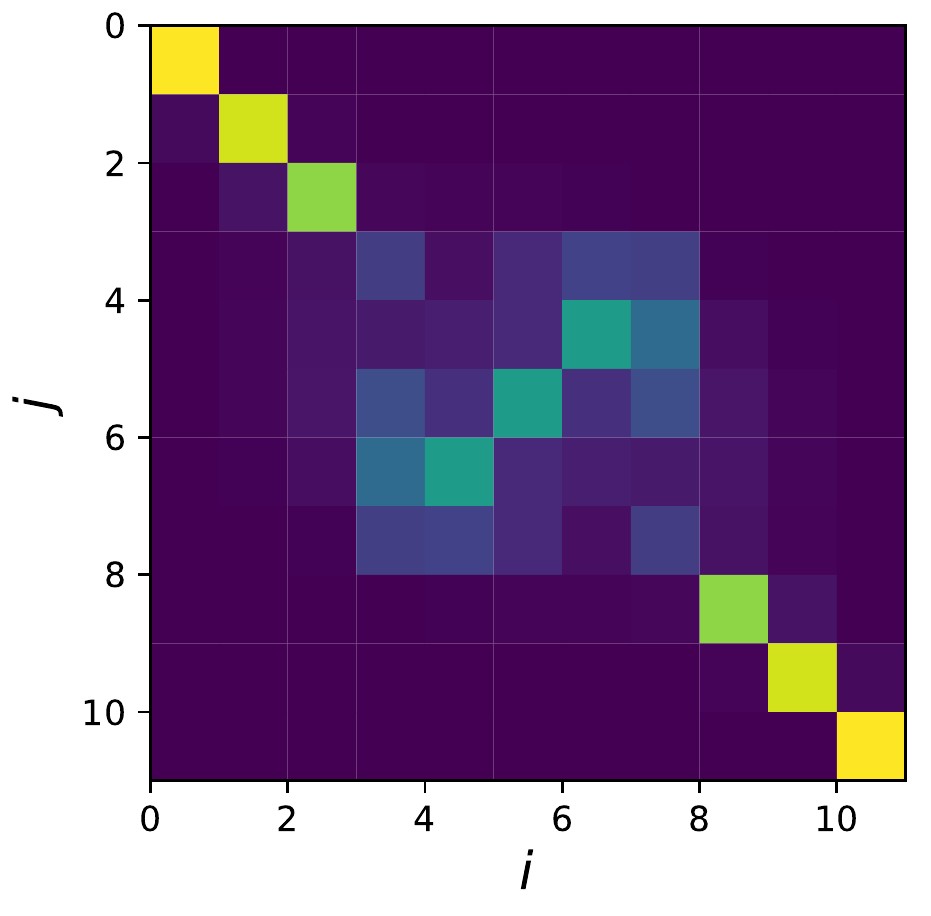}
    \label{fig:smin3IRAtau30r1}
  }
   \subfigure[\ \(s_{\min}=0.3, \tau =30, r=5\)]{
    \includegraphics[scale=0.4]{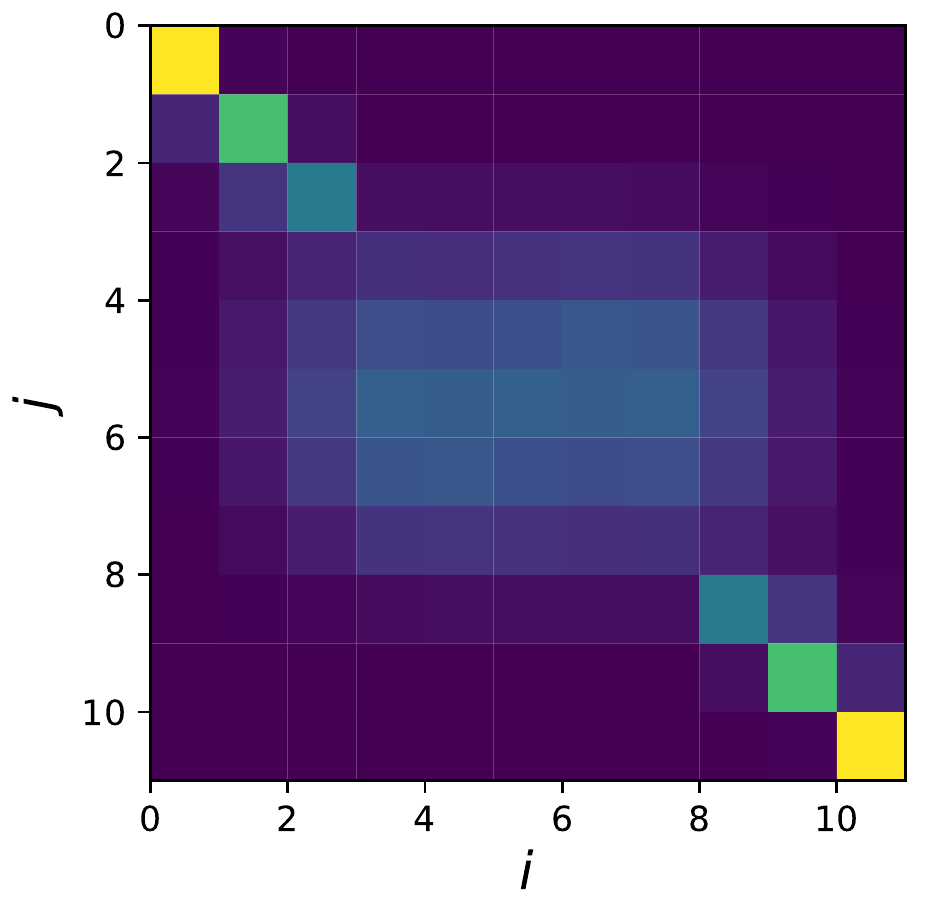}
    \label{fig:smin3IRAtau30r5}
  }
    \subfigure[\ \(s_{\min}=0.3, \tau =10, r=1\)]{
    \includegraphics[scale=0.4]{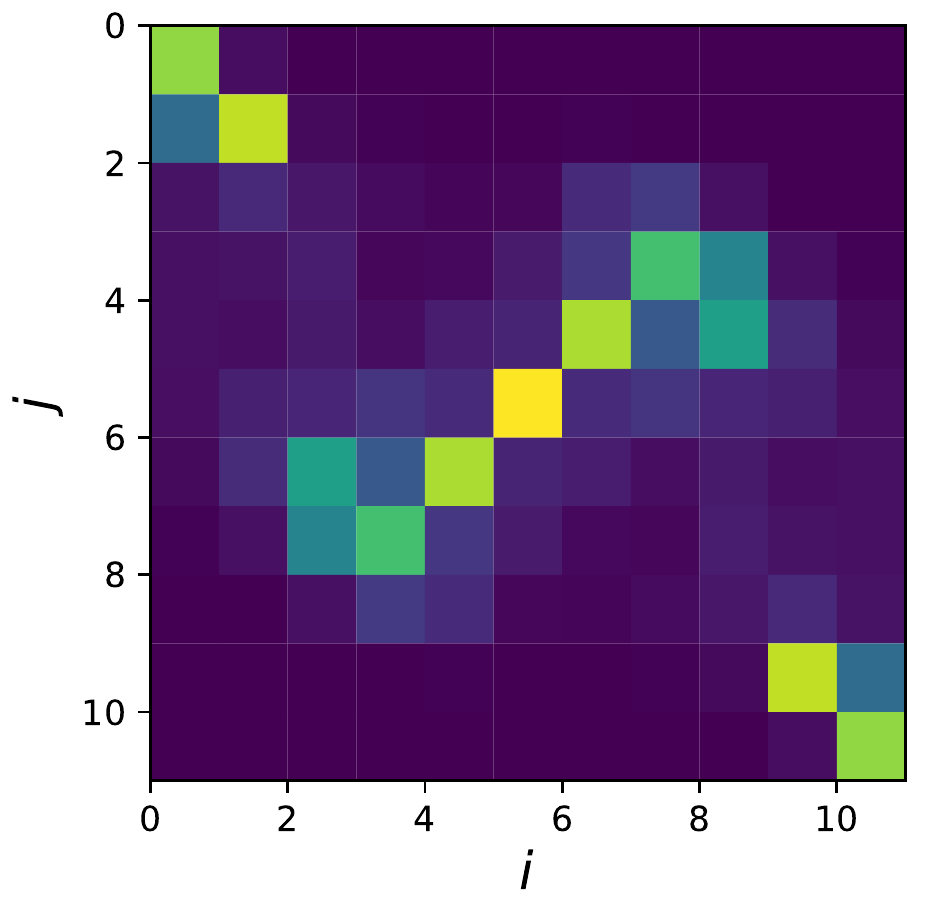}
    \label{fig:smin3IRAtau10r1}
  }
   \subfigure[\ \(s_{\min}=0.3, \tau =10, r=5\)]{
    \includegraphics[scale=0.4]{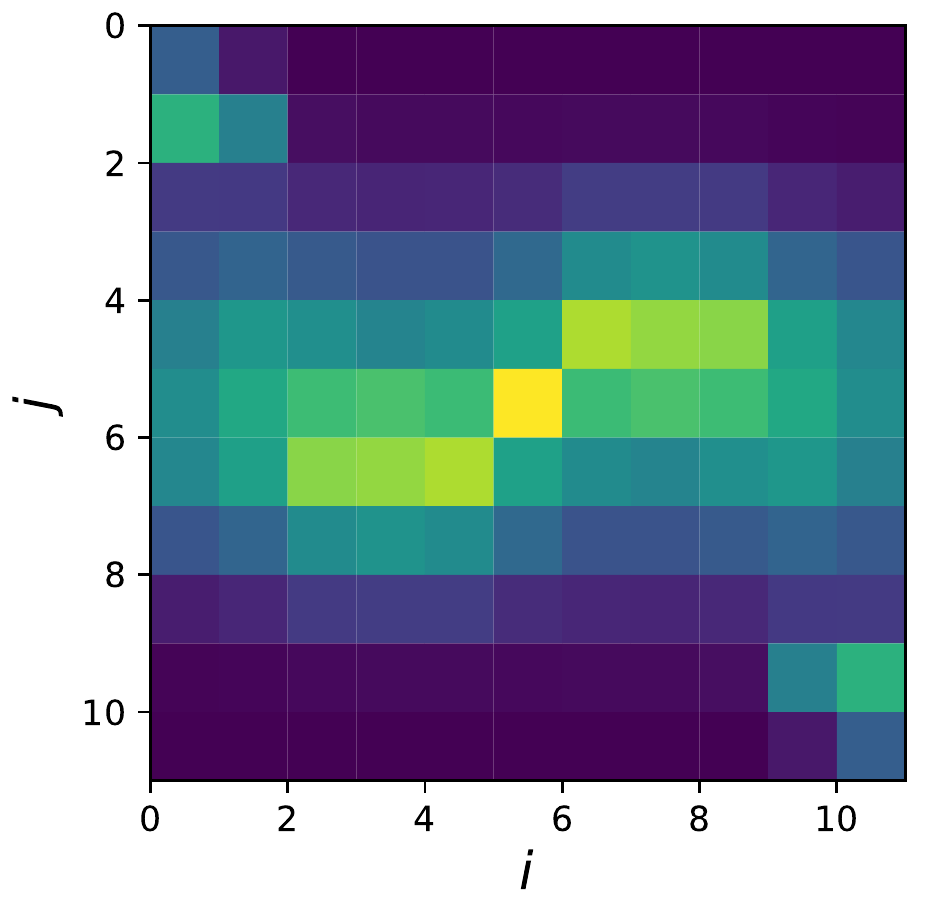}
    \label{fig:smin3IRAtau10r5}
  }
  
  \caption{Transition probabilities $(P^r)_{ji}$ from state $i$ to state $j$ for $p=3, ~N=10, ~s_{\min}=0.3$ after $r$ repetitions.
  Since $s_{\min}=0.3<s_c$, a phase transition occurs, accompanied by a relatively small energy gap between the ground state and the first excited state, which causes a diabatic transition to the first excited state, after which the state returns to the ground state at the second avoided crossing, which is represented as bright yellow in the top-left corner of panel (a).
  When started from an excited state $i>0$, the system does not reach the ground state. The reason is that the existence of many avoided crossings causes both upward and downward transitions, and does not unilaterally favor downward transitions toward the ground state.}
  \label{fig:IRAsmin03P}
\end{figure}

The density plots in Figs. \ref{fig:IRAsmin05P} and \ref{fig:IRAsmin03P} show the transition probabilities $(P^r)_{ji}$ from the initial state $i$ (horizontal axis) to the final state $j$ (vertical axis) in single cycles ($r=1$) and after multiple cycles ($r=3$ and $5$) with $s_{\min}=0.5$ (Fig.~\ref{fig:IRAsmin05P}) and  $s_{\min}=0.3$ (Fig.~\ref{fig:IRAsmin03P}).  For example, in Fig.~\ref{fig:IRAtau30r1}, if we start from the ground state $i=0$, the system reaches the same ground state  ($j=0$) with a large probability, as indicated by the yellow box, since the computation time $\tau=30$ is relatively long and close to adiabatic.  Other states along the column $i=0$ have lower probabilities (dark colors). Similarly, if the initial condition is an excited state $i>0$, the same state $j=i$ is reached with a high probability, resulting in a series of bright colors along the diagonal.  Repetition of IRA (panels (b) and (d)) or shorter computation time (panels (c) and (d)) results in scattering of the probabilities away from the diagonal. The situation is similar for $s_{\min}=0.3<s_c$ in Fig.~\ref{fig:IRAsmin03P}, but with a slightly different structure. In any case, the probability to reach the ground state at the end of the process, shown along the top row in each panel, remains low unless we start from the ground state itself, implying that IRA under coherent dynamics is unsuccessful in the present problem.

To what extent this is a special feature of the $p$-spin model, and whether IRA may work better in other cases, is an important topic for future work. Moreover, it is to be expected that thermal relaxation will reduce excited state occupation probabilities, so that an open system study of IRA is likely to find better performance for the finite (but low) temperature $p$-spin model than reported here.

\section{Conclusions}
Reverse quantum annealing protocols have generated much interest recently, as they may potentially overcome some of the obstacles that cause conventional forward quantum annealing to fail. In this work we presented a numerical study of the closed-system quantum dynamics of two types of reverse annealing, adiabatic reverse annealing (ARA) and iterated reverse annealing (IRA), for the $p$-spin model. For ARA, we demonstrated that the dynamical behavior of the system is consistent with its static phase diagram. The latter suggested that ARA may be able to exhibit an exponential speedup over QA for properly chosen initial conditions, by finding a path that avoids a first-order quantum phase transition that is encountered by QA. Our study confirms this expectation using the time-to-solution (TTS) measure: by establishing an optimal annealing time for each problem size we were able to extract the optimal TTS scaling, and show that it scales polynomially for ARA. At the same time we found that the lower bound for the TTS of QA is exponential. This is remarkable not only because ARA is thus a demonstrably better protocol than QA for at least one class of (admittedly trivial) optimization problems, but also since it lends credence to the predictions of the static analysis, which is often much simpler to perform than solving the dynamics. 

We also showed that ARA uses tunneling in the semiclassical potential to avoid a trap that the corresponding classical (zero temperature) spin vector dynamics cannot avoid. However, this trap is not an obstacle for the classical spin vector Monte Carlo algorithm, which uses thermal activation to hop over the corresponding potential barrier as long as the temperature is appropriately chosen.

In contrast to ARA, we found a negative result for IRA in the context of the $p$-spin model. In order for the IRA protocol to work and provide an enhanced probability of finding the ground state after multiple cycles, the probability distribution of the final state after a single cycle should shift toward lower energy states than the initial state. Unfortunately, this condition is not satisfied in the $p$-spin model as is clear from Figs.~\ref{fig:IRAsmin05P} and~\ref{fig:IRAsmin03P}. The generality of this conclusion is currently unclear, and it may be associated with the particular structure of the energy spectrum of the $p$-spin model, depicted in Fig.~\ref{fig:iRA_spect}. In particular, the gap structure is such that the $p$-spin model does not appear to lend itself to a diabatic cascade, wherein upward diabatic transitions are accompanied by downward transitions for properly tuned annealing schedules~\cite{Somma:2012kx,muthukrishnan_tunneling_2016,Brady:2017aa}. 
%although we should stay cautious since non-adiabatic dynamics is not completely described by the energy spectrum.  
However, our $p$-spin model result suggests the important lesson that there exist examples where IRA does not yield a ground state probability enhancement.

The present study concentrates on purely unitary quantum dynamics at zero temperature. In practice, real devices are exposed to an environment and are open quantum system with state transitions affected by thermal effects. In the D-Wave device, for example, thermal relaxation seems to lead to better performance as exemplified by the protocol of mid-anneal pausing \cite{Marshall2019}. It is an interesting future direction of research to test whether and how environmental effects change our conclusions, e.g., using methods such as those in Refs.~\cite{Passarelli2018,Passarelli2019}.

\section*{acknowledgement}
 This research is based upon work  supported by the Office of the Director of National Intelligence (ODNI), Intelligence Advanced Research Projects Activity (IARPA), via the U.S. Army Research Office contract W911NF-17-C-0050. The views and conclusions contained herein are those of the authors and should not be interpreted as necessarily representing the official policies or endorsements, either expressed or implied, of the ODNI, IARPA, or the U.S. Government. The U.S. Government is authorized to reproduce and distribute reprints for Governmental purposes notwithstanding any copyright annotation thereon. D.L. is additionally partially supported by DOE/HEP QuantISED program grant, QMLQCF (Quantum Machine Learning and Quantum Computation Frameworks for HEP), award number DE-SC0019227.

\bibliography{ReverseDynamics}

\begin{thebibliography}{39}%
\makeatletter
\providecommand \@ifxundefined [1]{%
 \@ifx{#1\undefined}
}%
\providecommand \@ifnum [1]{%
 \ifnum #1\expandafter \@firstoftwo
 \else \expandafter \@secondoftwo
 \fi
}%
\providecommand \@ifx [1]{%
 \ifx #1\expandafter \@firstoftwo
 \else \expandafter \@secondoftwo
 \fi
}%
\providecommand \natexlab [1]{#1}%
\providecommand \enquote  [1]{``#1''}%
\providecommand \bibnamefont  [1]{#1}%
\providecommand \bibfnamefont [1]{#1}%
\providecommand \citenamefont [1]{#1}%
\providecommand \href@noop [0]{\@secondoftwo}%
\providecommand \href [0]{\begingroup \@sanitize@url \@href}%
\providecommand \@href[1]{\@@startlink{#1}\@@href}%
\providecommand \@@href[1]{\endgroup#1\@@endlink}%
\providecommand \@sanitize@url [0]{\catcode `\\12\catcode `\$12\catcode
  `\&12\catcode `\#12\catcode `\^12\catcode `\_12\catcode `\%12\relax}%
\providecommand \@@startlink[1]{}%
\providecommand \@@endlink[0]{}%
\providecommand \url  [0]{\begingroup\@sanitize@url \@url }%
\providecommand \@url [1]{\endgroup\@href {#1}{\urlprefix }}%
\providecommand \urlprefix  [0]{URL }%
\providecommand \Eprint [0]{\href }%
\providecommand \doibase [0]{http://dx.doi.org/}%
\providecommand \selectlanguage [0]{\@gobble}%
\providecommand \bibinfo  [0]{\@secondoftwo}%
\providecommand \bibfield  [0]{\@secondoftwo}%
\providecommand \translation [1]{[#1]}%
\providecommand \BibitemOpen [0]{}%
\providecommand \bibitemStop [0]{}%
\providecommand \bibitemNoStop [0]{.\EOS\space}%
\providecommand \EOS [0]{\spacefactor3000\relax}%
\providecommand \BibitemShut  [1]{\csname bibitem#1\endcsname}%
\let\auto@bib@innerbib\@empty
%</preamble>
\bibitem [{\citenamefont {Kadowaki}\ and\ \citenamefont
  {Nishimori}(1998)}]{kadowaki_quantum_1998}%
  \BibitemOpen
  \bibfield  {author} {\bibinfo {author} {\bibfnamefont {T.}~\bibnamefont
  {Kadowaki}}\ and\ \bibinfo {author} {\bibfnamefont {H.}~\bibnamefont
  {Nishimori}},\ }\href {\doibase 10.1103/PhysRevE.58.5355} {\bibfield
  {journal} {\bibinfo  {journal} {Phys. Rev. E}\ }\textbf {\bibinfo {volume}
  {58}},\ \bibinfo {pages} {5355} (\bibinfo {year} {1998})}\BibitemShut
  {NoStop}%
\bibitem [{\citenamefont {Brooke}\ \emph {et~al.}(1999)\citenamefont {Brooke},
  \citenamefont {Bitko}, \citenamefont {F.}, \citenamefont {Rosenbaum},\ and\
  \citenamefont {Aeppli}}]{brooke_quantum_1999}%
  \BibitemOpen
  \bibfield  {author} {\bibinfo {author} {\bibfnamefont {J.}~\bibnamefont
  {Brooke}}, \bibinfo {author} {\bibfnamefont {D.}~\bibnamefont {Bitko}},
  \bibinfo {author} {\bibfnamefont {T.}~\bibnamefont {F.}}, \bibinfo {author}
  {\bibnamefont {Rosenbaum}}, \ and\ \bibinfo {author} {\bibfnamefont
  {G.}~\bibnamefont {Aeppli}},\ }\href {\doibase 10.1126/science.284.5415.779}
  {\bibfield  {journal} {\bibinfo  {journal} {Science}\ }\textbf {\bibinfo
  {volume} {284}},\ \bibinfo {pages} {779} (\bibinfo {year}
  {1999})}\BibitemShut {NoStop}%
\bibitem [{\citenamefont {Santoro}\ \emph {et~al.}(2002)\citenamefont
  {Santoro}, \citenamefont {Marto{\v n}{\'a}k}, \citenamefont {Tosatti},\ and\
  \citenamefont {Car}}]{santoro_theory_2002}%
  \BibitemOpen
  \bibfield  {author} {\bibinfo {author} {\bibfnamefont {G.~E.}\ \bibnamefont
  {Santoro}}, \bibinfo {author} {\bibfnamefont {R.}~\bibnamefont {Marto{\v
  n}{\'a}k}}, \bibinfo {author} {\bibfnamefont {E.}~\bibnamefont {Tosatti}}, \
  and\ \bibinfo {author} {\bibfnamefont {R.}~\bibnamefont {Car}},\ }\href
  {\doibase 10.1126/science.1068774} {\bibfield  {journal} {\bibinfo  {journal}
  {Science}\ }\textbf {\bibinfo {volume} {295}},\ \bibinfo {pages} {2427}
  (\bibinfo {year} {2002})}\BibitemShut {NoStop}%
\bibitem [{\citenamefont {Santoro}\ and\ \citenamefont
  {Tosatti}(2006)}]{santoro_optimization_2006}%
  \BibitemOpen
  \bibfield  {author} {\bibinfo {author} {\bibfnamefont {G.~E.}\ \bibnamefont
  {Santoro}}\ and\ \bibinfo {author} {\bibfnamefont {E.}~\bibnamefont
  {Tosatti}},\ }\href {https://doi.org/10.1088%2F0305-4470%2F39%2F36%2Fr01}
  {\bibfield  {journal} {\bibinfo  {journal} {J. Phys. A}\ }\textbf {\bibinfo
  {volume} {39}},\ \bibinfo {pages} {R393} (\bibinfo {year}
  {2006})}\BibitemShut {NoStop}%
\bibitem [{\citenamefont {Das}\ and\ \citenamefont
  {Chakrabarti}(2008)}]{das_colloquium:_2008}%
  \BibitemOpen
  \bibfield  {author} {\bibinfo {author} {\bibfnamefont {A.}~\bibnamefont
  {Das}}\ and\ \bibinfo {author} {\bibfnamefont {B.~K.}\ \bibnamefont
  {Chakrabarti}},\ }\href {\doibase 10.1103/RevModPhys.80.1061} {\bibfield
  {journal} {\bibinfo  {journal} {Rev. Mod. Phys.}\ }\textbf {\bibinfo {volume}
  {80}},\ \bibinfo {pages} {1061} (\bibinfo {year} {2008})}\BibitemShut
  {NoStop}%
\bibitem [{\citenamefont {Morita}\ and\ \citenamefont
  {Nishimori}(2008)}]{morita_mathematical_2008}%
  \BibitemOpen
  \bibfield  {author} {\bibinfo {author} {\bibfnamefont {S.}~\bibnamefont
  {Morita}}\ and\ \bibinfo {author} {\bibfnamefont {H.}~\bibnamefont
  {Nishimori}},\ }\href {\doibase 10.1063/1.2995837} {\bibfield  {journal}
  {\bibinfo  {journal} {J. Math. Phys.}\ }\textbf {\bibinfo {volume} {49}},\
  \bibinfo {pages} {125210} (\bibinfo {year} {2008})}\BibitemShut {NoStop}%
\bibitem [{\citenamefont {Farhi}\ \emph {et~al.}(2000)\citenamefont {Farhi},
  \citenamefont {Goldstone}, \citenamefont {Gutmann},\ and\ \citenamefont
  {Sipser}}]{farhi_quantum_2000}%
  \BibitemOpen
  \bibfield  {author} {\bibinfo {author} {\bibfnamefont {E.}~\bibnamefont
  {Farhi}}, \bibinfo {author} {\bibfnamefont {J.}~\bibnamefont {Goldstone}},
  \bibinfo {author} {\bibfnamefont {S.}~\bibnamefont {Gutmann}}, \ and\
  \bibinfo {author} {\bibfnamefont {M.}~\bibnamefont {Sipser}},\ }\href
  {http://arxiv.org/abs/quant-ph/0001106} {\bibfield  {journal} {\bibinfo
  {journal} {arXiv:quant-ph/0001106}\ } (\bibinfo {year} {2000})},\ \bibinfo
  {note} {arXiv: quant-ph/0001106}\BibitemShut {NoStop}%
\bibitem [{\citenamefont {Farhi}\ \emph {et~al.}(2001)\citenamefont {Farhi},
  \citenamefont {Goldstone}, \citenamefont {Gutmann}, \citenamefont {Lapan},
  \citenamefont {Lundgren},\ and\ \citenamefont {Preda}}]{farhi_quantum_2001}%
  \BibitemOpen
  \bibfield  {author} {\bibinfo {author} {\bibfnamefont {E.}~\bibnamefont
  {Farhi}}, \bibinfo {author} {\bibfnamefont {J.}~\bibnamefont {Goldstone}},
  \bibinfo {author} {\bibfnamefont {S.}~\bibnamefont {Gutmann}}, \bibinfo
  {author} {\bibfnamefont {J.}~\bibnamefont {Lapan}}, \bibinfo {author}
  {\bibfnamefont {A.}~\bibnamefont {Lundgren}}, \ and\ \bibinfo {author}
  {\bibfnamefont {D.}~\bibnamefont {Preda}},\ }\href {\doibase
  10.1126/science.1057726} {\bibfield  {journal} {\bibinfo  {journal}
  {Science}\ }\textbf {\bibinfo {volume} {292}},\ \bibinfo {pages} {472}
  (\bibinfo {year} {2001})}\BibitemShut {NoStop}%
\bibitem [{\citenamefont {Albash}\ and\ \citenamefont
  {Lidar}(2018{\natexlab{a}})}]{albash_adiabatic_2018}%
  \BibitemOpen
  \bibfield  {author} {\bibinfo {author} {\bibfnamefont {T.}~\bibnamefont
  {Albash}}\ and\ \bibinfo {author} {\bibfnamefont {D.~A.}\ \bibnamefont
  {Lidar}},\ }\href {\doibase 10.1103/RevModPhys.90.015002} {\bibfield
  {journal} {\bibinfo  {journal} {Rev. Mod. Phys.}\ }\textbf {\bibinfo {volume}
  {90}},\ \bibinfo {pages} {015002} (\bibinfo {year}
  {2018}{\natexlab{a}})}\BibitemShut {NoStop}%
\bibitem [{\citenamefont {Perdomo-Ortiz}\ \emph {et~al.}(2011)\citenamefont
  {Perdomo-Ortiz}, \citenamefont {Venegas-Andraca},\ and\ \citenamefont
  {Aspuru-Guzik}}]{perdomo-ortiz_study_2011}%
  \BibitemOpen
  \bibfield  {author} {\bibinfo {author} {\bibfnamefont {A.}~\bibnamefont
  {Perdomo-Ortiz}}, \bibinfo {author} {\bibfnamefont {S.~E.}\ \bibnamefont
  {Venegas-Andraca}}, \ and\ \bibinfo {author} {\bibfnamefont {A.}~\bibnamefont
  {Aspuru-Guzik}},\ }\href {\doibase 10.1007/s11128-010-0168-z} {\bibfield
  {journal} {\bibinfo  {journal} {Quantum Information Processing}\ }\textbf
  {\bibinfo {volume} {10}},\ \bibinfo {pages} {33} (\bibinfo {year}
  {2011})}\BibitemShut {NoStop}%
\bibitem [{\citenamefont {Chancellor}(2017)}]{Chancellor:2016ys}%
  \BibitemOpen
  \bibfield  {author} {\bibinfo {author} {\bibfnamefont {N.}~\bibnamefont
  {Chancellor}},\ }\href {\doibase 10.1088/1367-2630/aa59c4} {\bibfield
  {journal} {\bibinfo  {journal} {{New~J.~Phys.}}\ }\textbf {\bibinfo {volume}
  {19}},\ \bibinfo {pages} {023024} (\bibinfo {year} {2017})}\BibitemShut
  {NoStop}%
\bibitem [{\citenamefont {King}\ \emph {et~al.}(2018)\citenamefont {King},
  \citenamefont {Carrasquilla}, \citenamefont {Raymond}, \citenamefont
  {Ozfidan}, \citenamefont {Andriyash}, \citenamefont {Berkley}, \citenamefont
  {Reis}, \citenamefont {Lanting}, \citenamefont {Harris}, \citenamefont
  {Altomare}, \citenamefont {Boothby}, \citenamefont {Bunyk}, \citenamefont
  {Enderud}, \citenamefont {Fr{\'e}chette}, \citenamefont {Hoskinson},
  \citenamefont {Ladizinsky}, \citenamefont {Oh}, \citenamefont
  {Poulin-Lamarre}, \citenamefont {Rich}, \citenamefont {Sato}, \citenamefont
  {Smirnov}, \citenamefont {Swenson}, \citenamefont {Volkmann}, \citenamefont
  {Whittaker}, \citenamefont {Yao}, \citenamefont {Ladizinsky}, \citenamefont
  {Johnson}, \citenamefont {Hilton},\ and\ \citenamefont
  {Amin}}]{king_observation_2018}%
  \BibitemOpen
  \bibfield  {author} {\bibinfo {author} {\bibfnamefont {A.~D.}\ \bibnamefont
  {King}}, \bibinfo {author} {\bibfnamefont {J.}~\bibnamefont {Carrasquilla}},
  \bibinfo {author} {\bibfnamefont {J.}~\bibnamefont {Raymond}}, \bibinfo
  {author} {\bibfnamefont {I.}~\bibnamefont {Ozfidan}}, \bibinfo {author}
  {\bibfnamefont {E.}~\bibnamefont {Andriyash}}, \bibinfo {author}
  {\bibfnamefont {A.}~\bibnamefont {Berkley}}, \bibinfo {author} {\bibfnamefont
  {M.}~\bibnamefont {Reis}}, \bibinfo {author} {\bibfnamefont {T.}~\bibnamefont
  {Lanting}}, \bibinfo {author} {\bibfnamefont {R.}~\bibnamefont {Harris}},
  \bibinfo {author} {\bibfnamefont {F.}~\bibnamefont {Altomare}}, \bibinfo
  {author} {\bibfnamefont {K.}~\bibnamefont {Boothby}}, \bibinfo {author}
  {\bibfnamefont {P.~I.}\ \bibnamefont {Bunyk}}, \bibinfo {author}
  {\bibfnamefont {C.}~\bibnamefont {Enderud}}, \bibinfo {author} {\bibfnamefont
  {A.}~\bibnamefont {Fr{\'e}chette}}, \bibinfo {author} {\bibfnamefont
  {E.}~\bibnamefont {Hoskinson}}, \bibinfo {author} {\bibfnamefont
  {N.}~\bibnamefont {Ladizinsky}}, \bibinfo {author} {\bibfnamefont
  {T.}~\bibnamefont {Oh}}, \bibinfo {author} {\bibfnamefont {G.}~\bibnamefont
  {Poulin-Lamarre}}, \bibinfo {author} {\bibfnamefont {C.}~\bibnamefont
  {Rich}}, \bibinfo {author} {\bibfnamefont {Y.}~\bibnamefont {Sato}}, \bibinfo
  {author} {\bibfnamefont {A.~Y.}\ \bibnamefont {Smirnov}}, \bibinfo {author}
  {\bibfnamefont {L.~J.}\ \bibnamefont {Swenson}}, \bibinfo {author}
  {\bibfnamefont {M.~H.}\ \bibnamefont {Volkmann}}, \bibinfo {author}
  {\bibfnamefont {J.}~\bibnamefont {Whittaker}}, \bibinfo {author}
  {\bibfnamefont {J.}~\bibnamefont {Yao}}, \bibinfo {author} {\bibfnamefont
  {E.}~\bibnamefont {Ladizinsky}}, \bibinfo {author} {\bibfnamefont {M.~W.}\
  \bibnamefont {Johnson}}, \bibinfo {author} {\bibfnamefont {J.}~\bibnamefont
  {Hilton}}, \ and\ \bibinfo {author} {\bibfnamefont {M.~H.}\ \bibnamefont
  {Amin}},\ }\href {\doibase 10.1038/s41586-018-0410-x} {\bibfield  {journal}
  {\bibinfo  {journal} {Nature}\ }\textbf {\bibinfo {volume} {560}},\ \bibinfo
  {pages} {456} (\bibinfo {year} {2018})}\BibitemShut {NoStop}%
\bibitem [{\citenamefont {Ottaviani}\ and\ \citenamefont
  {Amendola}(2018)}]{Ottaviani2018}%
  \BibitemOpen
  \bibfield  {author} {\bibinfo {author} {\bibfnamefont {D.}~\bibnamefont
  {Ottaviani}}\ and\ \bibinfo {author} {\bibfnamefont {A.}~\bibnamefont
  {Amendola}},\ }\href {http://arxiv.org/abs/1808.08721} {\bibfield  {journal}
  {\bibinfo  {journal} {arXiv:1808.08721}\ } (\bibinfo {year} {2018})},\
  \Eprint {http://arxiv.org/abs/1808.08721} {arXiv:1808.08721} \BibitemShut
  {NoStop}%
\bibitem [{\citenamefont {Venturelli}\ and\ \citenamefont
  {Kondratyev}(2018)}]{Venturelli2018}%
  \BibitemOpen
  \bibfield  {author} {\bibinfo {author} {\bibfnamefont {D.}~\bibnamefont
  {Venturelli}}\ and\ \bibinfo {author} {\bibfnamefont {A.}~\bibnamefont
  {Kondratyev}},\ }\href {http://arxiv.org/abs/1810.08584} {\bibfield
  {journal} {\bibinfo  {journal} {arXiv:1810.08584}\ } (\bibinfo {year}
  {2018})},\ \Eprint {http://arxiv.org/abs/1810.08584} {arXiv:1810.08584}
  \BibitemShut {NoStop}%
\bibitem [{\citenamefont {Marshall}\ \emph {et~al.}(2019)\citenamefont
  {Marshall}, \citenamefont {Venturelli}, \citenamefont {Hen},\ and\
  \citenamefont {Rieffel}}]{Marshall2019}%
  \BibitemOpen
  \bibfield  {author} {\bibinfo {author} {\bibfnamefont {J.}~\bibnamefont
  {Marshall}}, \bibinfo {author} {\bibfnamefont {D.}~\bibnamefont
  {Venturelli}}, \bibinfo {author} {\bibfnamefont {I.}~\bibnamefont {Hen}}, \
  and\ \bibinfo {author} {\bibfnamefont {E.~G.}\ \bibnamefont {Rieffel}},\
  }\href {\doibase 10.1103/PhysRevApplied.11.044083} {\bibfield  {journal}
  {\bibinfo  {journal} {Phys. Rev. Applied}\ }\textbf {\bibinfo {volume}
  {11}},\ \bibinfo {pages} {044083} (\bibinfo {year} {2019})},\ \Eprint
  {http://arxiv.org/abs/1810.05881} {arXiv:1810.05881} \BibitemShut {NoStop}%
\bibitem [{\citenamefont {Ohkuwa}\ \emph {et~al.}(2018)\citenamefont {Ohkuwa},
  \citenamefont {Nishimori},\ and\ \citenamefont
  {Lidar}}]{ohkuwa_reverse_2018}%
  \BibitemOpen
  \bibfield  {author} {\bibinfo {author} {\bibfnamefont {M.}~\bibnamefont
  {Ohkuwa}}, \bibinfo {author} {\bibfnamefont {H.}~\bibnamefont {Nishimori}}, \
  and\ \bibinfo {author} {\bibfnamefont {D.~A.}\ \bibnamefont {Lidar}},\ }\href
  {\doibase 10.1103/PhysRevA.98.022314} {\bibfield  {journal} {\bibinfo
  {journal} {Phys. Rev. A}\ }\textbf {\bibinfo {volume} {98}} (\bibinfo {year}
  {2018}),\ 10.1103/PhysRevA.98.022314}\BibitemShut {NoStop}%
\bibitem [{\citenamefont {Pfeuty}(1970)}]{Pfeuty1970}%
  \BibitemOpen
  \bibfield  {author} {\bibinfo {author} {\bibfnamefont {P.}~\bibnamefont
  {Pfeuty}},\ }\href@noop {} {\bibfield  {journal} {\bibinfo  {journal} {Annals
  of Physics}\ }\textbf {\bibinfo {volume} {90}},\ \bibinfo {pages} {79}
  (\bibinfo {year} {1970})}\BibitemShut {NoStop}%
\bibitem [{\citenamefont {Laumann}\ \emph {et~al.}(2012)\citenamefont
  {Laumann}, \citenamefont {Moessner}, \citenamefont {Scardicchio},\ and\
  \citenamefont {Sondhi}}]{Laumann:2012hs}%
  \BibitemOpen
  \bibfield  {author} {\bibinfo {author} {\bibfnamefont {C.~R.}\ \bibnamefont
  {Laumann}}, \bibinfo {author} {\bibfnamefont {R.}~\bibnamefont {Moessner}},
  \bibinfo {author} {\bibfnamefont {A.}~\bibnamefont {Scardicchio}}, \ and\
  \bibinfo {author} {\bibfnamefont {S.~L.}\ \bibnamefont {Sondhi}},\ }\href
  {http://link.aps.org/doi/10.1103/PhysRevLett.109.030502} {\bibfield
  {journal} {\bibinfo  {journal} {Phys. Rev. Lett.}\ }\textbf {\bibinfo
  {volume} {109}},\ \bibinfo {pages} {030502} (\bibinfo {year}
  {2012})}\BibitemShut {NoStop}%
\bibitem [{\citenamefont {Tsuda}\ \emph {et~al.}(2013)\citenamefont {Tsuda},
  \citenamefont {Yamanaka},\ and\ \citenamefont {Nishimori}}]{Tsuda2013}%
  \BibitemOpen
  \bibfield  {author} {\bibinfo {author} {\bibfnamefont {J.}~\bibnamefont
  {Tsuda}}, \bibinfo {author} {\bibfnamefont {Y.}~\bibnamefont {Yamanaka}}, \
  and\ \bibinfo {author} {\bibfnamefont {H.}~\bibnamefont {Nishimori}},\ }\href
  {\doibase 10.7566/JPSJ.82.114004} {\bibfield  {journal} {\bibinfo  {journal}
  {J. Phys. Soc. Jpn.}\ }\textbf {\bibinfo {volume} {82}},\ \bibinfo {pages}
  {114004} (\bibinfo {year} {2013})}\BibitemShut {NoStop}%
\bibitem [{\citenamefont {Kato}(1950)}]{Kato:50}%
  \BibitemOpen
  \bibfield  {author} {\bibinfo {author} {\bibfnamefont {T.}~\bibnamefont
  {Kato}},\ }\href {\doibase 10.1143/JPSJ.5.435} {\bibfield  {journal}
  {\bibinfo  {journal} {J. Phys. Soc. Jpn.}\ }\textbf {\bibinfo {volume} {5}},\
  \bibinfo {pages} {435} (\bibinfo {year} {1950})}\BibitemShut {NoStop}%
\bibitem [{\citenamefont {Jansen}\ \emph {et~al.}(2007)\citenamefont {Jansen},
  \citenamefont {Ruskai},\ and\ \citenamefont {Seiler}}]{Jansen:07}%
  \BibitemOpen
  \bibfield  {author} {\bibinfo {author} {\bibfnamefont {S.}~\bibnamefont
  {Jansen}}, \bibinfo {author} {\bibfnamefont {M.-B.}\ \bibnamefont {Ruskai}},
  \ and\ \bibinfo {author} {\bibfnamefont {R.}~\bibnamefont {Seiler}},\ }\href
  {http://scitation.aip.org/content/aip/journal/jmp/48/10/10.1063/1.2798382}
  {\bibfield  {journal} {\bibinfo  {journal} {J. Math. Phys.}\ }\textbf
  {\bibinfo {volume} {48}},\ \bibinfo {pages} {102111} (\bibinfo {year}
  {2007})}\BibitemShut {NoStop}%
\bibitem [{\citenamefont {Lidar}\ \emph {et~al.}(2009)\citenamefont {Lidar},
  \citenamefont {Rezakhani},\ and\ \citenamefont {Hamma}}]{lidar:102106}%
  \BibitemOpen
  \bibfield  {author} {\bibinfo {author} {\bibfnamefont {D.~A.}\ \bibnamefont
  {Lidar}}, \bibinfo {author} {\bibfnamefont {A.~T.}\ \bibnamefont
  {Rezakhani}}, \ and\ \bibinfo {author} {\bibfnamefont {A.}~\bibnamefont
  {Hamma}},\ }\href
  {http://scitation.aip.org/content/aip/journal/jmp/50/10/10.1063/1.3236685}
  {\bibfield  {journal} {\bibinfo  {journal} {J. Math. Phys.}\ }\textbf
  {\bibinfo {volume} {50}},\ \bibinfo {pages} {102106} (\bibinfo {year}
  {2009})}\BibitemShut {NoStop}%
\bibitem [{\citenamefont {Amin}(2015)}]{Amin:2015qf}%
  \BibitemOpen
  \bibfield  {author} {\bibinfo {author} {\bibfnamefont {M.~H.}\ \bibnamefont
  {Amin}},\ }\href {https://link.aps.org/doi/10.1103/PhysRevA.92.052323}
  {\bibfield  {journal} {\bibinfo  {journal} {Phys. Rev. A}\ }\textbf {\bibinfo
  {volume} {92}},\ \bibinfo {pages} {052323} (\bibinfo {year}
  {2015})}\BibitemShut {NoStop}%
\bibitem [{\citenamefont {Munoz-Bauza}\ \emph {et~al.}(2019)\citenamefont
  {Munoz-Bauza}, \citenamefont {Chen},\ and\ \citenamefont
  {Lidar}}]{Munoz-Bauza:2019aa}%
  \BibitemOpen
  \bibfield  {author} {\bibinfo {author} {\bibfnamefont {H.}~\bibnamefont
  {Munoz-Bauza}}, \bibinfo {author} {\bibfnamefont {H.}~\bibnamefont {Chen}}, \
  and\ \bibinfo {author} {\bibfnamefont {D.}~\bibnamefont {Lidar}},\ }\href
  {\doibase 10.1038/s41534-019-0160-0} {\bibfield  {journal} {\bibinfo
  {journal} {npj Quantum Information}\ }\textbf {\bibinfo {volume} {5}},\
  \bibinfo {pages} {2} (\bibinfo {year} {2019})}\BibitemShut {NoStop}%
\bibitem [{\citenamefont {Karanikolas}\ and\ \citenamefont
  {Kawabata}(2018)}]{Karanikolas:2018aa}%
  \BibitemOpen
  \bibfield  {author} {\bibinfo {author} {\bibfnamefont {V.}~\bibnamefont
  {Karanikolas}}\ and\ \bibinfo {author} {\bibfnamefont {S.}~\bibnamefont
  {Kawabata}},\ }\href {https://arxiv.org/abs/1806.08517} {\bibfield  {journal}
  {\bibinfo  {journal} {arXiv preprint arXiv:1806.08517}\ } (\bibinfo {year}
  {2018})}\BibitemShut {NoStop}%
\bibitem [{hga()}]{hgain}%
  \BibitemOpen
  \href@noop {} {}\bibinfo {note}
  {\url{https://docs.dwavesys.com/docs/latest/c_fd_hg.html}}\BibitemShut
  {NoStop}%
\bibitem [{\citenamefont {J{\"{o}}rg}\ \emph {et~al.}(2010)\citenamefont
  {J{\"{o}}rg}, \citenamefont {Krzakala}, \citenamefont {Kurchan},
  \citenamefont {Maggs},\ and\ \citenamefont {Pujos}}]{Jorg2010}%
  \BibitemOpen
  \bibfield  {author} {\bibinfo {author} {\bibfnamefont {T.}~\bibnamefont
  {J{\"{o}}rg}}, \bibinfo {author} {\bibfnamefont {F.}~\bibnamefont
  {Krzakala}}, \bibinfo {author} {\bibfnamefont {J.}~\bibnamefont {Kurchan}},
  \bibinfo {author} {\bibfnamefont {a.~C.}\ \bibnamefont {Maggs}}, \ and\
  \bibinfo {author} {\bibfnamefont {J.}~\bibnamefont {Pujos}},\ }\href
  {\doibase 10.1209/0295-5075/89/40004} {\bibfield  {journal} {\bibinfo
  {journal} {EPL}\ }\textbf {\bibinfo {volume} {89}},\ \bibinfo {pages} {40004}
  (\bibinfo {year} {2010})}\BibitemShut {NoStop}%
\bibitem [{\citenamefont {R{\o}nnow}\ \emph {et~al.}(2014)\citenamefont
  {R{\o}nnow}, \citenamefont {Wang}, \citenamefont {Job}, \citenamefont
  {Boixo}, \citenamefont {Isakov}, \citenamefont {Wecker}, \citenamefont
  {Martinis}, \citenamefont {Lidar},\ and\ \citenamefont {Troyer}}]{speedup}%
  \BibitemOpen
  \bibfield  {author} {\bibinfo {author} {\bibfnamefont {T.~F.}\ \bibnamefont
  {R{\o}nnow}}, \bibinfo {author} {\bibfnamefont {Z.}~\bibnamefont {Wang}},
  \bibinfo {author} {\bibfnamefont {J.}~\bibnamefont {Job}}, \bibinfo {author}
  {\bibfnamefont {S.}~\bibnamefont {Boixo}}, \bibinfo {author} {\bibfnamefont
  {S.~V.}\ \bibnamefont {Isakov}}, \bibinfo {author} {\bibfnamefont
  {D.}~\bibnamefont {Wecker}}, \bibinfo {author} {\bibfnamefont {J.~M.}\
  \bibnamefont {Martinis}}, \bibinfo {author} {\bibfnamefont {D.~A.}\
  \bibnamefont {Lidar}}, \ and\ \bibinfo {author} {\bibfnamefont
  {M.}~\bibnamefont {Troyer}},\ }\href
  {http://science.sciencemag.org/content/345/6195/420} {\bibfield  {journal}
  {\bibinfo  {journal} {Science}\ }\textbf {\bibinfo {volume} {345}},\ \bibinfo
  {pages} {420} (\bibinfo {year} {2014})}\BibitemShut {NoStop}%
\bibitem [{\citenamefont {Albash}\ and\ \citenamefont
  {Lidar}(2018{\natexlab{b}})}]{Albash:2017aa}%
  \BibitemOpen
  \bibfield  {author} {\bibinfo {author} {\bibfnamefont {T.}~\bibnamefont
  {Albash}}\ and\ \bibinfo {author} {\bibfnamefont {D.~A.}\ \bibnamefont
  {Lidar}},\ }\href {\doibase 10.1103/PhysRevX.8.031016} {\bibfield  {journal}
  {\bibinfo  {journal} {Phys. Rev. X}\ }\textbf {\bibinfo {volume} {8}},\
  \bibinfo {pages} {031016} (\bibinfo {year} {2018}{\natexlab{b}})}\BibitemShut
  {NoStop}%
\bibitem [{\citenamefont {Hen}\ \emph {et~al.}(2015)\citenamefont {Hen},
  \citenamefont {Job}, \citenamefont {Albash}, \citenamefont {Ronnow},
  \citenamefont {Troyer},\ and\ \citenamefont {Lidar}}]{Hen:2015rt}%
  \BibitemOpen
  \bibfield  {author} {\bibinfo {author} {\bibfnamefont {I.}~\bibnamefont
  {Hen}}, \bibinfo {author} {\bibfnamefont {J.}~\bibnamefont {Job}}, \bibinfo
  {author} {\bibfnamefont {T.}~\bibnamefont {Albash}}, \bibinfo {author}
  {\bibfnamefont {T.~F.}\ \bibnamefont {Ronnow}}, \bibinfo {author}
  {\bibfnamefont {M.}~\bibnamefont {Troyer}}, \ and\ \bibinfo {author}
  {\bibfnamefont {D.~A.}\ \bibnamefont {Lidar}},\ }\href
  {http://link.aps.org/doi/10.1103/PhysRevA.92.042325} {\bibfield  {journal}
  {\bibinfo  {journal} {{Phys. Rev. A}}\ }\textbf {\bibinfo {volume} {92}},\
  \bibinfo {pages} {042325} (\bibinfo {year} {2015})}\BibitemShut {NoStop}%
\bibitem [{\citenamefont {Smolin}\ and\ \citenamefont {Smith}(2014)}]{Smolin}%
  \BibitemOpen
  \bibfield  {author} {\bibinfo {author} {\bibfnamefont {J.~A.}\ \bibnamefont
  {Smolin}}\ and\ \bibinfo {author} {\bibfnamefont {G.}~\bibnamefont {Smith}},\
  }\href {http://journal.frontiersin.org/article/10.3389/fphy.2014.00052}
  {\bibfield  {journal} {\bibinfo  {journal} {Frontiers in Physics}\ }\textbf
  {\bibinfo {volume} {2}},\ \bibinfo {pages} {52} (\bibinfo {year}
  {2014})}\BibitemShut {NoStop}%
\bibitem [{\citenamefont {Albash}\ \emph {et~al.}(2015)\citenamefont {Albash},
  \citenamefont {R{\o}nnow}, \citenamefont {Troyer},\ and\ \citenamefont
  {Lidar}}]{albash_reexamining_2015}%
  \BibitemOpen
  \bibfield  {author} {\bibinfo {author} {\bibfnamefont {T.}~\bibnamefont
  {Albash}}, \bibinfo {author} {\bibfnamefont {T.}~\bibnamefont {R{\o}nnow}},
  \bibinfo {author} {\bibfnamefont {M.}~\bibnamefont {Troyer}}, \ and\ \bibinfo
  {author} {\bibfnamefont {D.}~\bibnamefont {Lidar}},\ }\href {\doibase
  10.1140/epjst/e2015-02346-0} {\bibfield  {journal} {\bibinfo  {journal} {The
  European Physical Journal Special Topics}\ }\textbf {\bibinfo {volume}
  {224}},\ \bibinfo {pages} {111} (\bibinfo {year} {2015})}\BibitemShut
  {NoStop}%
\bibitem [{\citenamefont {Shin}\ \emph {et~al.}(2014)\citenamefont {Shin},
  \citenamefont {Smith}, \citenamefont {Smolin},\ and\ \citenamefont
  {Vazirani}}]{Shin2014}%
  \BibitemOpen
  \bibfield  {author} {\bibinfo {author} {\bibfnamefont {S.~W.}\ \bibnamefont
  {Shin}}, \bibinfo {author} {\bibfnamefont {G.}~\bibnamefont {Smith}},
  \bibinfo {author} {\bibfnamefont {J.~A.}\ \bibnamefont {Smolin}}, \ and\
  \bibinfo {author} {\bibfnamefont {U.}~\bibnamefont {Vazirani}},\ }\href
  {http://arxiv.org/abs/1401.7087} {\bibfield  {journal} {\bibinfo  {journal}
  {arXiv:1401.7087}\ } (\bibinfo {year} {2014})},\ \Eprint
  {http://arxiv.org/abs/1401.7087} {arXiv:1401.7087} \BibitemShut {NoStop}%
\bibitem [{\citenamefont {Muthukrishnan}\ \emph {et~al.}(2016)\citenamefont
  {Muthukrishnan}, \citenamefont {Albash},\ and\ \citenamefont
  {Lidar}}]{muthukrishnan_tunneling_2016}%
  \BibitemOpen
  \bibfield  {author} {\bibinfo {author} {\bibfnamefont {S.}~\bibnamefont
  {Muthukrishnan}}, \bibinfo {author} {\bibfnamefont {T.}~\bibnamefont
  {Albash}}, \ and\ \bibinfo {author} {\bibfnamefont {D.~A.}\ \bibnamefont
  {Lidar}},\ }\href {\doibase 10.1103/PhysRevX.6.031010} {\bibfield  {journal}
  {\bibinfo  {journal} {Phys. Rev. X}\ }\textbf {\bibinfo {volume} {6}},\
  \bibinfo {pages} {031010} (\bibinfo {year} {2016})}\BibitemShut {NoStop}%
\bibitem [{\citenamefont {Seki}\ and\ \citenamefont
  {Nishimori}(2012)}]{seki_quantum_2012}%
  \BibitemOpen
  \bibfield  {author} {\bibinfo {author} {\bibfnamefont {Y.}~\bibnamefont
  {Seki}}\ and\ \bibinfo {author} {\bibfnamefont {H.}~\bibnamefont
  {Nishimori}},\ }\href {\doibase 10.1103/PhysRevE.85.051112} {\bibfield
  {journal} {\bibinfo  {journal} {Phys. Rev. E}\ }\textbf {\bibinfo {volume}
  {85}},\ \bibinfo {pages} {051112} (\bibinfo {year} {2012})}\BibitemShut
  {NoStop}%
\bibitem [{\citenamefont {Somma}\ \emph {et~al.}(2012)\citenamefont {Somma},
  \citenamefont {Nagaj},\ and\ \citenamefont {Kieferov{\'a}}}]{Somma:2012kx}%
  \BibitemOpen
  \bibfield  {author} {\bibinfo {author} {\bibfnamefont {R.~D.}\ \bibnamefont
  {Somma}}, \bibinfo {author} {\bibfnamefont {D.}~\bibnamefont {Nagaj}}, \ and\
  \bibinfo {author} {\bibfnamefont {M.}~\bibnamefont {Kieferov{\'a}}},\ }\href
  {http://link.aps.org/doi/10.1103/PhysRevLett.109.050501} {\bibfield
  {journal} {\bibinfo  {journal} {Phys. Rev. Lett.}\ }\textbf {\bibinfo
  {volume} {109}},\ \bibinfo {pages} {050501} (\bibinfo {year}
  {2012})}\BibitemShut {NoStop}%
\bibitem [{\citenamefont {Brady}\ and\ \citenamefont {van
  Dam}(2017)}]{Brady:2017aa}%
  \BibitemOpen
  \bibfield  {author} {\bibinfo {author} {\bibfnamefont {L.~T.}\ \bibnamefont
  {Brady}}\ and\ \bibinfo {author} {\bibfnamefont {W.}~\bibnamefont {van
  Dam}},\ }\href {\doibase 10.1103/PhysRevA.95.032335} {\bibfield  {journal}
  {\bibinfo  {journal} {Physical Review A}\ }\textbf {\bibinfo {volume} {95}},\
  \bibinfo {pages} {032335} (\bibinfo {year} {2017})}\BibitemShut {NoStop}%
\bibitem [{\citenamefont {Passarelli}\ \emph {et~al.}(2018)\citenamefont
  {Passarelli}, \citenamefont {{De Filippis}}, \citenamefont {Cataudella},\
  and\ \citenamefont {Lucignano}}]{Passarelli2018}%
  \BibitemOpen
  \bibfield  {author} {\bibinfo {author} {\bibfnamefont {G.}~\bibnamefont
  {Passarelli}}, \bibinfo {author} {\bibfnamefont {G.}~\bibnamefont {{De
  Filippis}}}, \bibinfo {author} {\bibfnamefont {V.}~\bibnamefont
  {Cataudella}}, \ and\ \bibinfo {author} {\bibfnamefont {P.}~\bibnamefont
  {Lucignano}},\ }\href {\doibase 10.1103/PhysRevA.97.022319} {\bibfield
  {journal} {\bibinfo  {journal} {Phys. Rev. A}\ }\textbf {\bibinfo {volume}
  {97}},\ \bibinfo {pages} {022319} (\bibinfo {year} {2018})},\ \Eprint
  {http://arxiv.org/abs/1801.07491} {arXiv:1801.07491} \BibitemShut {NoStop}%
\bibitem [{\citenamefont {Passarelli}\ \emph {et~al.}(2019)\citenamefont
  {Passarelli}, \citenamefont {Cataudella},\ and\ \citenamefont
  {Lucignano}}]{Passarelli2019}%
  \BibitemOpen
  \bibfield  {author} {\bibinfo {author} {\bibfnamefont {G.}~\bibnamefont
  {Passarelli}}, \bibinfo {author} {\bibfnamefont {V.}~\bibnamefont
  {Cataudella}}, \ and\ \bibinfo {author} {\bibfnamefont {P.}~\bibnamefont
  {Lucignano}},\ }\href {http://arxiv.org/abs/1902.06788} {\bibfield  {journal}
  {\bibinfo  {journal} {arXiv:1902.06788}\ } (\bibinfo {year} {2019})},\
  \Eprint {http://arxiv.org/abs/1902.06788} {arXiv:1902.06788} \BibitemShut
  {NoStop}%
\end{thebibliography}%

\end{document}